\documentclass[12pt,thmsa,a4paper]{article}
\usepackage{amssymb}
\usepackage{sw20elba}

\input tcilatex
\QQQ{Language}{
British English
}

\begin{document}

\title{\textbf{Principles of Discrete Time Mechanics:}\\
$\mathbf{IV:}$\textbf{\ The Dirac equation, particles and oscillons}}
\author{Keith Norton and George Jaroszkiewicz \\
Department of Mathematics, University of Nottingham\\
University Park, Nottingham NG7 2RD, UK}
\date{\today }
\maketitle

\begin{abstract}
\textit{We apply the principles of discrete time mechanics discussed in
earlier papers to the first and second quantised Dirac equation. We use the
Schwinger action principle to find the anticommutation relations of the
Dirac field and of the particle creation operators in the theory. We find
new solutions to the discrete time Dirac equation, referred to as }\emph{%
oscillons} \textit{on account of their extraordinary behaviour. Their
principal characteristic is that they oscillate with a period twice that of
the fundamental time interval }$T$ \textit{of our theory.\thinspace Although
these solutions can be associated with definite charge, linear momentum and
spin, such objects should not be observable as particles in the continuous
time limit. We find that for non-zero }$T$ \textit{they correspond to states
with negative squared norm in Hilbert space. However they are an integral
part of the discrete time Dirac field and should play a role in particle
interactions analogous to the role of longitudinal photons in conventional
quantum electrodynamics.}
\end{abstract}

\section{Introduction}

Throughout this paper we shall use the acronyms \textit{CT }to denote\textit{%
\ continuous time} and \textit{DT} to denote \textit{discrete time}. The
symbol $T$ is used to denote our fundamental interval of time, which is
assumed positive. The term \textit{CT limit} and the symbol $\stackunder{T}{%
\rightarrow }$ will refer to the taking of $T$ to zero and some integer $n$
to infinity, such that $nT\stackunder{T}{\rightarrow }t$, where $t$ is
ordinary continuous co-ordinate time, assuming that this limit makes sense.
This will frequently be the case, but it should always be kept in mind that 
\textit{DT} mechanics is inherently richer in its range of possibilities
than \textit{CT} mechanics and includes trajectories for which such a limit
is meaningless. This is particularly the case for the \textit{oscillon}
solutions which we shall discuss in the case of the \textit{DT} Dirac
equation.

This paper is the fourth in a series devoted to the construction of \textit{%
DT} classical and quantum mechanics from first principles. This series of
papers examines two questions: first, whether it is possible or consistent
to rewrite all the laws of physics using discrete time rather than
continuous time, and second, if such a programme were viable and self
consistent, what would be the advantages (and perhaps disadvantages) of such
a discrete time formulation. Throughout this series we invert the standard
view of discrete time. Rather than regarding it as a convenient
approximation to continuous time, suitable for say computer simulation, we
suppose from the outset that our ordinary conception of time is itself an
approximation, albeit an extraordinarily useful one, and that deep down
there is a scale at which temporal phenomena occur in a discrete way and not
in a differentiable or even continuous way. An analogy with water is useful
here. For a vast range of phenomena, water may be sensibly modelled as a
continuum, but sooner or later a molecular approach must be used.

Our philosophy imposes a rigid discipline on our work. Our principles of
discrete time mechanics must be as well specified as those of continuous
time mechanics. Our laws of motion must be exact, and our invariants of the
motion cannot be approximately conserved but must be exactly conserved
modulo those equations of motion. We should avoid introducing ad-hoc fixes
for problems when they occur, but rely solely on the basic starting
assumptions. In this respect we have found that our basic principles have
guided us well, so that once a calculation has been started, we have found
virtually no freedom of choice thereafter. What has emerged is a consistent
theory of mechanics, both classical and quantum, which is genuinely
different from continuous time mechanics, but which has sufficient overlap
with it in so many ways as to make further investigation highly desirable.

A question of some delicacy for \textit{DT }mechanics is the issue of
Lorentz covariance. Our \textit{DT} mechanics is not manifestly Lorentz
covariant. Fortunately, it turns out from our studies of scalar and Dirac
field propagators that Lorentz covariance and the Poincar\'{e} algebra
appear to be broken in \textit{DT }mechanics only at the order $T^2$ level
so that our field theoretic scattering amplitudes are expected to have an
expansion in powers of $T$ such that the zeroth order terms are Lorentz
symmetric. From our perspective, then, Lorentz covariance emerges as an
approximate symmetry of the complete mechanics, rather like isotopic
symmetry is an approximate symmetry of nuclear physics. If as we imagine $T$
is of the order of the Planck time or less, then there should be no problem
in confronting current particle data, because special relativity has not
been tested to anywhere near Planck scales.

This raises an obvious question; why bother to discretise time if all that
it does is to produce amplitudes which virtually duplicate conventional
field theory? Our answer is that there are known problems with conventional
formulations of field theory based on continuous time, such as divergences
in Feynman diagrams, and our study is an exploration of an alternative
approach to mechanics which might alleviate some of these problems.

Our first paper, referred to as \textit{Principles I} $\cite{J&N-I},$
introduced basic principles for the temporal discretisation of \textit{CT}
classical and quantum particle mechanics. We have recently found a paper by
Khorrami \cite{K-94}, who discussed various similar topics and which is
substantially in agreement with our results. Our second paper, referred to
as \textit{Principles II} $\cite{J&N-II},$ extended and applied our
principles to classical field theory, including the Schr\"{o}dinger
equation, the Klein-Gordon equation, Maxwell's equations, gauge invariant
electrodynamics and the classical Dirac field. The third paper, \textit{%
Principles III,} \cite{J&N-III} tackled scalar quantum field theory and
there we discussed the construction of Feynman rules for \textit{DT }scalar
field theory. In that paper we found that otherwise hard or point-like
vertices in \textit{CT} $\varphi ^3$ Feynman diagrams are replaced by
softened vertices in \textit{DT} $\varphi ^{3\text{ }}$ theory, and that
there is a conserved quantity analogous to energy in particle scattering
processes, even though there is not a Hamiltonian in the theory. These
papers should be consulted for further explanation of our notation,
methodology, and motivation. In the present paper we apply the techniques of
the previous papers\emph{\ }to a discussion of the \textit{DT} formulation
of the quantised Dirac equation. We shall use natural units throughout,
where $c=\hbar =1.$

The next section is a review of the fundamentally important \textit{DT}
harmonic oscillator, which provides the basic template for all our field
theory propagators and generates the momentum space cutoff predicted in the
free particle spectrum. Then we turn to the Dirac equation in one time and
zero space dimensions. This provides a toy model which allows us to explore
the novelties of \textit{DT} field theory. This model generates the normal
spectrum of particle and anti-particle states, and also entirely novel
solutions which, by virtue of their temporal behaviour, we call \textit{%
oscillons} and \textit{anti-oscillons.} The fundamentally bizarre property
of oscillons is that they oscillate in phase with a period twice that of the
fundamental time $T$. We may readily appreciate that, in the temporal limit $%
T\rightarrow 0$, these oscillon solutions cannot be accommodated within
conventional \textit{CT} mechanics. If $T$ is extremely small but not zero,
we may think of these oscillons as ghost like particles, capable of carrying
momentum, spin and charge, but not really a normal form of matter. It is an
important result that our quantisation procedure, based on the \textit{DT}
Schwinger action principle, tells us that such single particle oscillon
states, were they to be created, would be represented by state vectors
having a negative inner product amongst themselves. These states would be
regarded as unphysical, according to the usual interpretation of such
vectors in state space.

We then turn to the \textit{DT} Dirac equation in one time and three spatial
dimensions. The results are much as with the $1+0$ toy model discussed
previously, but with some obvious differences. Now we can consider linear
momentum and spin. We show how to construct the corresponding conserved
variables using the \textit{Maeda-Noether} theorem discussed in \textit{%
Principles I }and \textit{II}. We note that in \textit{DT} we can examine
field commutators and anti-commutators in fine detail, and there are one or
two surprises with the Dirac field anti-commutators, related to the
existence of oscillon solutions. We show also how to couple the \textit{DT}
Dirac equation to the \textit{DT} Maxwell potentials in a gauge invariant
way, and discuss the equations of motion and the conserved electric charge.
Finally, we finish with a statement of the \textit{DT} reduction formulae
for the particle and anti-particle \textit{in} and \textit{out} states, in
preparation for application to \textit{DT }quantum electrodynamics, to be
studied in subsequent papers of this series.

\section{The Scalar field in 1+0 dimensions}

In this section we review the inhomogeneous \textit{DT }harmonic oscillator
equation, which serves as the basis for all \textit{DT} free field
equations, including the Klein-Gordon equation, the Dirac equation, and
Maxwell's equations.

We showed in \textit{Principles I }that\textit{\ }the classical \textit{DT}
equation of motion for an harmonic oscillator in the presence of a source $j$
is given by the Cadzow equation \cite{J&N-I,CADZOW.70} 
\begin{equation}
\beta x_{n+1}-2\alpha x_n+\beta x_{n-1}\stackunder{c}{=}Tj_n,\;\;\;\;\;\beta
\neq 0,  \label{B1}
\end{equation}
where we use the symbol $\stackunder{c}{=}$ to denote any equality holding
on a true or classical trajectory, and $\alpha $ and $\beta $ are constants
which are determined by the \textit{CT }Lagrangian from which the \textit{DT 
}system function is constructed. This equation can be written in the form 
\begin{equation}
\left( U_n-2\eta +U_n^{-1}\right) x_n\stackunder{c}{=}J_n,  \label{B2}
\end{equation}
where $\eta \equiv \alpha /\beta $, $J_n\equiv Tj_n/\beta $ and $U_n$ is the
classical step operator with the property $U_nf_n=f_{n+1}$ for any variable $%
f$ indexed by integer $n$. Equation $\left( \ref{B2}\right) $ is a second
order difference equation. We recall from our previous papers that for this
particular equation there are two important dynamical regimes, called the 
\textit{elliptic} and \textit{hyperbolic} regimes, corresponding to $\eta
^2<1$ and $\eta ^2>1$ respectively.\textit{\ }These are separated by the 
\textit{parabolic barrier, }$\eta ^2=1$. Oscillatory behaviour occurs in the
elliptic regime, which in field theories corresponds to the region where
physical particle states occur. Singularities in propagators occur at the
parabolic barrier in \textit{DT }field theory, and avoiding these requires
some careful discussion.

To solve the equation of motion $\left( \ref{B1}\right) $, first define the
weighted differences 
\begin{equation}
\Delta x_n\equiv x_n-\mu x_{n-1},  \label{B3}
\end{equation}
and suppose that they satisfy the first-order difference equation 
\begin{equation}
\Delta x_{n+1}\stackunder{c}{=}\lambda \Delta x_n+J_n  \label{B4}
\end{equation}
for each value of $n$ considered and for some complex-valued constants $%
\lambda $ and $\mu $. Then we find 
\begin{equation}
\lambda +\mu =2\eta ,\;\;\;\lambda \mu =1,  \label{B5}
\end{equation}
which have solutions 
\begin{equation}
\mu =\eta \pm \sqrt{\eta ^2-1},\;\;\;\lambda =\eta \mp \sqrt{\eta ^2-1}.
\label{B6}
\end{equation}
In the case of forwards propagation, we find 
\begin{eqnarray}
x_N\stackunder{c}{=}P^{N-n}x_n-P^{N-n-1}x_{n-1}+P^{N-n-1}J_{n\;\;\;\;\;\;\;}
&&  \nonumber \\
+...+P^1J_{N-2}+P^0J_{N-1},\;\;\;\;\;\;\;\;\;\;N &>&n,  \label{B7}
\end{eqnarray}
whereas for backwards propagation we find 
\begin{eqnarray}
x_M\stackunder{c}{=}-P^{n-M-1}x_{n+1}+P^{n-M}x_n+P^{n-M-1}J_n &&  \nonumber
\\
+...+P^1J_{M+2}+P^0J_{M+1},\;\;\;\;\;\;\;M &<&n,  \label{B8}
\end{eqnarray}
where the polynomials $P^n\equiv P^n(\lambda ,\mu )$ are defined by 
\begin{equation}
P^n(\lambda ,\mu )\equiv \lambda ^n+\lambda ^{n-1}\mu +...+\lambda \mu
^{n-1}+\mu ^n,  \label{B9}
\end{equation}
where $P^1\left( \lambda ,\mu \right) \equiv \lambda +\mu $ and $P^0\left(
\lambda ,\mu \right) \equiv 1.$ From $\left( \ref{B7}\right) $ and $\left( 
\ref{B8}\right) $ we find 
\begin{eqnarray}
P^{n-M-1}x_N+P^{N-n-1}x_M\stackunder{c}{=}
&&(P^{n-M-1}P^{N-n}+P^{N-n-1}P^{n-M}-P^1P^{n-M-1}P^{N-n-1})x_n  \nonumber \\
&&-P^{n-M-1}P^{N-n-1}J_n+P^{n-M-1}\sum_{i=n}^{N-1}P^{N-1-i}J_i \\
&&\;\;\;\;\;+P^{N-n-1}\sum_{i=n}^{M+1}P^{i-M-1}J_i,\;\;\;\;\;M<n<N.
\label{B11}
\end{eqnarray}
This is the desired relationship between the particle's position in the past
(assumed known at time $MT<nT$), the particle's position in the future
(assumed known at time $NT>nT$), and the particle's position at the present
time $nT$. This equation incorporates Feynman (causal) propagation and the
effect of the source terms and is an exact result.

The next step is to look at the scattering limit $N\rightarrow \infty $, $%
M\rightarrow -\infty .$ Potential singularities at the parabolic barrier are
avoided by taking the prescription $\eta \rightarrow \eta +i\varepsilon
,\;\varepsilon >0,$ which corresponds to the Feynman $m\rightarrow $ $%
m-i\varepsilon $ prescription in field theory, where $\epsilon $ is
infinitesimal and positive. It is convenient to introduce the parameter $%
\theta $, related to $\eta $ by\ $\eta =\cos \theta $ and permitting it to
take values in the complex $\theta $ plane only on the contour $\theta =u-iv$%
, where 
\begin{eqnarray}
0 &\leq &u<\pi ,\;\;v=\varepsilon >0,\;\;\;\;\;\;\;\;(\text{the elliptic
regime)}  \nonumber \\
u &=&\pi ,\;\;\;\;\;\;\;\varepsilon \leq v<\infty \;\;\;\left( \text{the
hyperbolic regime}\right) .
\end{eqnarray}
By writing $\eta \equiv x+iy$ we see 
\begin{equation}
\left( \frac x{\cosh v}\right) ^2+\left( \frac y{\sinh v}\right) ^2=1.
\end{equation}
Points in the elliptic regime lie on the upper half of an ellipse in the
complex $\eta $ plane enclosing the points $\eta =\pm 1$, which shows that
the parabolic barrier is avoided.

Turning to the $P^n$ functions, we use the fact that $\lambda \mu =1$ to
find 
\begin{equation}
P^n\left( \lambda ,\mu \right) =\frac{\mu ^{n+1}-\mu ^{-n-1}}{\mu -\mu ^{-1}}%
,\;\;\;\;\;\mu ^2\neq 1.
\end{equation}
Then taking$\;\mu =e^{i\theta }=e^{iu+v}$ we find 
\begin{equation}
\lim_{n\rightarrow \infty }\frac{P^n}{e^{n\left( iu+v\right) }}=\frac{%
e^{iu+v}}{e^{iu+v}-e^{-iu-v}}.
\end{equation}
Now taking the scattering limit $N=-M\rightarrow \infty $ in $\left( \ref
{B11}\right) $ we find 
\begin{equation}
x_n\sim x_n^{\left( 0\right) }-T\sum_{m=-\infty }^\infty \Delta _F^{n-m}j_m,
\end{equation}
where $x_n^{\left( 0\right) }$ satisfies the source free equation and the 
\textit{DT} indexed Feynman propagator $\Delta _F^n$ satisfies the equation 
\begin{equation}
\left( U_n-2\eta +U_n^{-1}\right) \Delta _F^n=-\Gamma \delta
_n,\;\;\;\;\;\;\Gamma \equiv \beta ^{-1},  \label{bose}
\end{equation}
where $\delta _n$ is the \textit{DT} Kronecker delta defined in $\left( \ref
{delta}\right) $. Solving $\left( \ref{bose}\right) $ we find 
\begin{equation}
\Delta _F^n=\frac{\Gamma e^{-i|n|\theta }}{2i\sin \theta },\;\;\;\;\;\theta
=u-iv,  \label{sta}
\end{equation}
which holds for the elliptic and hyperbolic regimes when we take into
account the range of values the parameter $\theta $ could take.

Turning to particle theory and the Klein-Gordon equation, we saw in \cite
{J&N-II} that the various constants in the above equations have the
following parametrisations:

\begin{equation}
\Gamma \equiv \beta ^{-1}=\frac{6T}{6+T^2E^2},\;\;\;\eta =\frac{6-2T^2E^2}{%
6+T^2E^2},\;\;\;
\end{equation}
where $E\equiv \sqrt{\mathbf{p.p}+m^2}$ is by definition the energy of the
particle and $\mathbf{p}$ is its linear momentum. Now we define the
transformed propagator 
\begin{equation}
\tilde{\Delta}_F\left( \mathbf{p},\Theta \right) \equiv T\sum_{n=-\infty
}^\infty e^{in\Theta }\Delta _F^n\left( \mathbf{p}\right) ,
\end{equation}
which satisfies the equation 
\begin{equation}
2\left( \cos \Theta -\eta \right) \tilde{\Delta}_F\left( \mathbf{p},\Theta
\right) =-T\Gamma .
\end{equation}
Thence we find 
\begin{equation}
\tilde{\Delta}_F\left( \mathbf{p},\Theta \right) =\frac{-T\Gamma }{2\left(
\cos \Theta -\eta -i\epsilon \right) },  \label{pprop}
\end{equation}
using the Feynman $+i\varepsilon $ prescription for avoiding the
singularities. In the above we have assumed $\Theta $ is real. We may define
the propagator for complex values of $\Theta $ by analytic continuation of $%
\left( \ref{pprop}\right) .$ If now we introduce the variable $p_0$ related
to $\Theta $ by the rule 
\begin{equation}
\cos \Theta \equiv \frac{6-2p_0^2T^2}{6+p_0^2T^2},\;\;\;sign\left( \Theta
\right) =sign\left( p_0\right) ,  \label{bparam}
\end{equation}
then we find 
\begin{equation}
\tilde{\Delta}_F\left( \mathbf{p},\Theta \right) =\frac 1{\left( p_0^2-%
\mathbf{p}^2-m^2+i\epsilon \right) }+\frac{T^2p_0^2}{6\left( p_0^2-\mathbf{p}%
^2-m^2+i\epsilon \right) },  \label{bprop}
\end{equation}
an exact result. From this we see the emergence of Lorentz symmetry as an
approximate symmetry of the mechanics. If $p_0$ in the above is taken to
represent the same thing in \textit{DT} as it does in \textit{CT} special
relativity, i.e., the zeroth component of a four-vector, then the first term
on the right hand side is clearly a Lorentz scalar, corresponding to the
standard bosonic propagator of \textit{CT} field theory. The second term on
the right hand side is not Lorentz invariant but is down on the first term
by a factor proportional to $T^2.$ If as we expect $T$ represents an
extremely small time scale, such as the Planck time or less, then it is
clear that the second term in $\left( \ref{bprop}\right) $ will be so much
smaller in its effects than the first in general and so may be neglected in
practice. This is why we expect our \textit{DT} quantum field theory to be
extremely well approximated by Lorentz covariant \textit{CT} field theory in
most situations.

We can use the propagator and the \textit{DT} Schwinger action principle to
calculate the ground state expectation of \textit{DT} time-ordered products
of field operators, and hence extract the free field commutation and
anticommutation relations. This was done in \textit{Paper III }for the
scalar field and we shall do the same for the Dirac equation in this paper.

\section{The quantised Dirac particle in 1+0 dimensions}

In this section we discuss the quantised \textit{DT }Dirac particle in one
time and zero spatial dimensions. This serves as the prototype for the Dirac
field studied in the next section. We will use the Schwinger source function
technique to obtain the ground state functional in the presence of the
sources and from that we are able to extract the ground state expectation
values of various anticommutators.

\subsection{Classical equations}

We start with the \textit{CT} theory. Our dynamical variables are $\psi
,\psi ^{+}$, which have two fermionic (anticommuting) degrees of freedom
each: 
\begin{equation}
\psi =\left[ 
\begin{array}{c}
\psi _1 \\ 
\psi _2
\end{array}
\right] ,\;\;\;\;\;\psi ^{+}=\left[ 
\begin{array}{cc}
\psi _1^{*} & \psi _2^{*}
\end{array}
\right] ,\;\;\;\;\;
\end{equation}
with $\;\bar{\psi}\equiv \psi ^{+}\gamma ^0,$ where 
\begin{equation}
\gamma ^0=\left[ 
\begin{array}{cc}
1 & 0 \\ 
0 & -1
\end{array}
\right] .
\end{equation}
Then the \textit{CT }Lagrangian is 
\begin{eqnarray}
L &=&\frac{_1}{^2}i\bar{\psi}\gamma ^0\overrightarrow{\partial _t}\psi -%
\frac{_1}{^2}i\bar{\psi}\overleftarrow{\partial _t}\gamma ^0\psi -m\bar{\psi}%
\psi  \nonumber \\
&=&\frac{_1}{^2}i\psi ^{+}\dot{\psi}-\frac{_1}{^2}i\dot{\psi}^{+}\psi -m\psi
^{+}\gamma ^0\psi ,
\end{eqnarray}
where we assume the mass $m$ is non-zero$.$\ Following \textit{Principles I}
and \textit{II }we obtain the \textit{DT }system function $F^n$ from the 
\textit{CT} Lagrangian by considering the virtual paths 
\begin{eqnarray}
\tilde{\psi}_n &=&\lambda \psi _{n+1}+\bar{\lambda}\psi _n,  \nonumber \\
\tilde{\psi}_n^{+} &=&\lambda \psi _{n+1}^{+}+\bar{\lambda}\psi
_n^{+},\;\;\;\;\;\bar{\lambda}\equiv 1-\lambda ,
\end{eqnarray}
and integrating $TL(\tilde{\psi}_n,\,\tilde{\psi}_n^{+})$ from $\lambda =0$
to $\lambda =1$. With this prescription we find the time derivatives turn
into differences. The result is the system function 
\begin{eqnarray}
F^n &=&\frac{_1}{^2}i\left\{ \psi _n^{+}\psi _{n+1}-\psi _{n+1}^{+}\psi
_n\right\}  \nonumber \\
&&-\frac \kappa 6\left\{ 2\psi _{n+1}^{+}\gamma ^0\psi _{n+1}+\psi
_n^{+}\gamma ^0\psi _{n+1}+\psi _{n+1}^{+}\gamma ^0\psi _n+2\psi
_n^{+}\gamma ^0\psi _n\right\} ,
\end{eqnarray}
where $\kappa \equiv mT$.

Next we add the Schwinger sources, which are taken as infinitesimal external
disturbances to the system. This may be done in a number of ways. Because we
will be interested in vacuum expectation values when the sources are
switched off, it does not really matter what we choose to add and how. Our
choice turns out to be the most convenient. We define the system function $%
F^n\left[ \eta \right] $ in the presence of external (i.e. non dynamical)
fermionic sources $\eta $, $\bar{\eta}$ to be given by 
\begin{equation}
F^n\left[ \eta \right] \equiv F^n+\frac{{}_1}{{}^2}T\left\{ \bar{\eta}_n\psi
_n+\bar{\eta}_{n+1}\psi _{n+1}+\bar{\psi}_n\eta _n+\bar{\psi}_{n+1}\eta
_{n+1}\right\}
\end{equation}
and then the Cadzow equation of motion 
\begin{equation}
\frac \partial {\partial \psi _n^{+}}\left\{ F^n\left[ \eta \right]
+F^{n-1}\left[ \eta \right] \right\} \stackunder{c}{=}0
\end{equation}
gives 
\begin{equation}
\frac{i\gamma ^0}{2T}\left( \psi _{n+1}-\psi _{n-1}\right) -\frac m6\left(
\psi _{n+1}+4\psi _n+\psi _{n-1}\right) \stackunder{c}{=}-\eta _n,
\label{orig}
\end{equation}
or 
\begin{equation}
\omega ^{+}\psi _{n+1}+4\kappa \psi _n+\omega \psi _{n-1}\stackunder{c}{=}%
6T\eta _n,  \label{Deq}
\end{equation}
where $\omega \equiv \kappa +3i\gamma ^0,$ with a similar equation for the
conjugate variable $\psi ^{+}.$

Now define the non-singular matrix 
\begin{equation}
\hat{\omega}\equiv \frac \omega {|\omega |}=e^{i\xi \gamma ^0}=\cos \xi
+i\sin \xi \gamma ^0=\left[ 
\begin{array}{cc}
e^{i\xi } & 0 \\ 
0 & e^{-i\xi }
\end{array}
\right] ,\;\;\;
\end{equation}
where $|\omega |\equiv \sqrt{9+\kappa ^2}$ and the phase angle $\xi $
satisfies $\left( \ref{zeta}\right) .$ Then the equation of motion $\left( 
\ref{Deq}\right) $ becomes 
\begin{equation}
\left( \hat{\omega}^{-1}U_n-2\eta +\hat{\omega}U_n^{-1}\right) \psi _n=\frac{%
6T}{|\omega |}\eta _n,
\end{equation}
where $\eta \equiv -2\kappa /|\omega |.$ To solve this equation we first
simplify the Dirac space dependence by the non-singular transformation $\psi
_n\equiv \hat{\omega}^n\phi _n$ and then the equation of motion becomes 
\begin{equation}
\;\;\;\left( U_n-2\eta +U_n^{-1}\right) \phi _n\stackunder{c}{=}\frac{6T}{%
|\omega |}\hat{\omega}^{-n}\eta _n.
\end{equation}
Using the results discussed in \S $2$ we obtain the formal scattering
solution 
\begin{equation}
\phi _n=\phi _n^{\left( 0\right) }-T\sum_{m=-\infty }^\infty \Delta _F^{n-m}%
\hat{\omega}^{-m}\eta _m,
\end{equation}
where $\phi _n^{\left( 0\right) }$ is a solution of the source free equation
and the bosonic propagator satisfies equation $\left( \ref{bose}\right) .$
This solution incorporates Feynman scattering boundary conditions.
Transforming back to the original fields, we find 
\begin{equation}
\psi _n=\psi _n^{\left( 0\right) }-T\sum_{m=-\infty }^\infty S_F^{n-m}\eta
_m,
\end{equation}
where 
\begin{eqnarray}
S_F^n &\equiv &\Delta _F^n\hat{\omega}^n=\frac 6{2i|\omega |\sin \theta
}e^{-i|n|\theta }\hat{\omega}^n \\
&=&\frac{6e^{-i|n|\theta }}{2i|\omega |\sin \theta }\left\{ \cos n\xi +i\sin
n\xi \;\gamma ^0\right\}
\end{eqnarray}
is the required \textit{DT }Dirac propagator. It satisfies the equation 
\begin{equation}
\left( \hat{\omega}^{-1}U_n-2\eta +\hat{\omega}U_n^{-1}\right) S_F^n=-\frac
6{|\omega |}\delta _n,
\end{equation}
which is equivalent to 
\begin{equation}
\left\{ \frac{i\gamma ^0}{2T}\left( U_n-U_n^{-1}\right) -\frac m6\left(
U_n+4+U_n^{-1}\right) \right\} S_F^n=\frac{\delta _n}T.
\end{equation}
There are two important angles in this theory, given by 
\begin{eqnarray}
\cos \theta &=&\frac{-2\kappa }{\sqrt{9+\kappa ^2}},\;\;\;\sin \theta =\frac{%
\sqrt{9-3\kappa ^2}}{\sqrt{9+\kappa ^2}}  \label{theta} \\
\cos \xi &=&\frac \kappa {\sqrt{9+\kappa ^2}},\;\;\;\sin \xi =\frac 3{\sqrt{%
9+\kappa ^2}}  \label{zeta}
\end{eqnarray}
where we assume we are in the elliptic regime $9-3\kappa ^2>0$. It is
particular linear combinations of these angles which characterise the
solutions to the \textit{DT }Dirac equation, as we shall now show. If we
define the angles $\delta \equiv \theta -\xi ,\;\sigma \equiv \theta +\xi $
then we find 
\begin{eqnarray}
S_F^0 &=&\frac 3{i\sqrt{9-3\kappa ^2}},  \nonumber \\
S_F^n &=&\frac 3{i\sqrt{9-3\kappa ^2}}\left[ 
\begin{array}{cc}
e^{-in\delta } & 0 \\ 
0 & e^{-in\sigma }
\end{array}
\right] ,\;\;\;n>0 \\
S_F^n &=&\frac 3{i\sqrt{9-3\kappa ^2}}\left[ 
\begin{array}{cc}
e^{in\sigma } & 0 \\ 
0 & \;\;e^{in\delta }
\end{array}
\right] ,\;\;\;n<0.  \nonumber
\end{eqnarray}

The significance of the angles $\delta $ and $\xi $ is the following.
Assuming the fundamental parameter $T$ is vanishingly small, a Taylor
expansion about $\kappa =0$ gives 
\begin{eqnarray}
\delta &\equiv &\theta -\xi \simeq \kappa +\frac 1{180}\kappa ^5+O\left(
\kappa ^7\right) ,  \nonumber \\
\sigma &\equiv &\theta +\xi \simeq \pi +\frac 13\kappa +O\left( \kappa
^3\right) .
\end{eqnarray}
It turns out that the phase $\delta $ is associated with physical particles
in the \textit{CT }limit $T\rightarrow 0$, whereas $\sigma $ is associated
with wave-functions which change sign more-or-less once during each
fundamental timestep $T$. This highly oscillatory behaviour is quite
different to any behaviour encountered in \textit{CT} mechanics and leads us
to coin the term \textit{oscillon }for solutions to the equations which
behave in such a bizarre way. In our theory there will be anti-oscillons as
well as oscillons, in the same way there are anti-particles as well as
particles.

The reason oscillon solutions occur is not hard to understand. Our \textit{DT%
} Dirac equation $\left( \ref{orig}\right) $ is, unlike the \textit{CT}
Dirac equation, manifestly second-order in nature. Such an equation will
normally have two solutions. However, there are twice as many degrees of
freedom when we take into account the Dirac space dimensions in the above
theory, so we find a total of four solutions to the \textit{DT} Dirac
equation. These correspond to \textit{particle, anti-particle, oscillon, }%
and \textit{anti-oscillon} solutions. In \textit{Paper II} we discussed the 
\textit{DT} Schr\"{o}dinger equation in exactly analogous terms, except we
did not have a Dirac space structure. So in that example we found particle
wavefunction and oscillon wavefunction solutions but no anti-particle or
anti-oscillon solutions. We argued there that the oscillon wave-function
solutions should not be accessible under ordinary conditions. This is
supported by the result we discuss below, that quantum oscillon particle
states have a negative norm and so are properly regarded as unphysical.

\subsection{Quantisation}

Having found the propagator the next step is to extract information from it
about the field operators. We note that unlike \emph{CT} mechanics, where we
are usually able to impose canonical commutation or anticommutation
relations directly, there is more of a problem in \emph{DT} mechanics. This
is because the concept of canonical conjugate momentum becomes close to
being redundant in our theory. Fortunately, the Schwinger source functional
approach provides a powerful way of finding commutation or anticommutation
relations which are consistent with the quantum dynamics.

Given a \emph{DT} action sum operator $\hat{A}^{NM}\left[ \eta \right] ,$
then for evolution from state $|\beta ,M\rangle $ at time $TM$ to $|\alpha
,N\rangle $ at time $TN$ $\left( N>M\right) $ in the presence of external
sources, our \emph{DT }Schwinger action principle is just 
\begin{equation}
\delta \langle \alpha ,N|\beta ,M\rangle _\eta =i\langle \alpha ,N|\delta 
\hat{A}^{NM}\left[ \eta \right] |\beta ,M\rangle _\eta ,\;\;\;N>M.
\end{equation}
Here the action sum operator is given by 
\begin{eqnarray}
\hat{A}^{NM}\left[ \eta \right] &\equiv &\sum_{n=M}^{N-1}\hat{F}^n\left[
\eta \right]  \nonumber \\
&=&\sum_{n=M}^{N-1}\{\hat{F}^n+\frac{_1}{^2}T[\bar{\eta}_{n+1}\hat{\psi}%
_{n+1}+\bar{\eta}_n\hat{\psi}_n+\bar{\psi}_{n+1}\eta _{n+1}+\bar{\psi}_n\eta
_n]\}.
\end{eqnarray}
Then we find 
\begin{eqnarray}
\frac{-i}T\frac \partial {\partial \bar{\eta}_n}\langle \alpha ,N|\beta
,M\rangle _\eta &=&\langle \alpha ,N|\hat{\psi}_n|\beta ,M\rangle _\eta
,\;\;\;N>n>M  \nonumber \\
\frac iT\frac \partial {\partial \eta _n}\langle \alpha ,N|\beta ,M\rangle
_\eta &=&\langle \alpha ,N|\bar{\psi}_n|\beta ,M\rangle _\eta ,\;\;\;N>n>M
\label{aa} \\
\frac{-i}T\frac \partial {\partial \bar{\eta}_N}\langle \alpha ,N|\beta
,M\rangle _\eta &=&\frac{_1}{^2}\langle \alpha ,N|\hat{\psi}_N|\beta
,M\rangle _\eta ,  \nonumber \\
\frac iT\frac \partial {\partial \eta _N}\langle \alpha ,N|\beta ,M\rangle
_\eta &=&\frac{_1}{^2}\langle \alpha ,N|\bar{\psi}_N|\beta ,M\rangle _\eta ,
\\
\frac{-i}T\frac \partial {\partial \bar{\eta}_M}\langle \alpha ,N|\beta
,M\rangle _\eta &=&\frac{_1}{^2}\langle \alpha ,N|\hat{\psi}_M|\beta
,M\rangle _\eta ,  \nonumber \\
\frac iT\frac \partial {\partial \eta _M}\langle \alpha ,N|\beta ,M\rangle
_\eta &=&\frac{_1}{^2}\langle \alpha ,N|\bar{\psi}_M|\beta ,M\rangle _\eta .
\label{bb}
\end{eqnarray}
We may use these results to work out \textit{DT} ground state time ordered
products. For fermionic fields we note for example 
\begin{eqnarray}
\langle 0|\tilde{T}\hat{\psi}_n^{+}\hat{\psi}_m|0\rangle &\equiv &\langle 0|%
\hat{\psi}_n^{+}\hat{\psi}_m|0\rangle \Theta _{n-m}+\frac{_1}{^2}\langle 0|[%
\hat{\psi}_n^{+},\hat{\psi}_n]|0\rangle \delta _{n-m} \\
&&-\langle 0|\hat{\psi}_m\hat{\psi}_n^{+}|0\rangle \Theta _{m-n},  \nonumber
\end{eqnarray}
where $\tilde{T}$ denotes the \textit{DT} time ordering operator discussed
in \cite{J&N-I,J&N-II,J&N-III}, and $\Theta _{n\text{ }}$and $\delta _n$ are
the \textit{DT} step and $\delta $ functions respectively, with the
properties 
\begin{eqnarray}
\Theta _{n\text{ }} &=&1,\;\;\;n>0,\;\;\;\;\;\;\;\;\;\;\delta _n=1,\;\;\;n=0,
\nonumber \\
&=&0,\;\;\;n<1,\;\;\;\;\;\;\;\;\;\;\;\;\;\;=0,\;\;\;n\neq 0.  \label{delta}
\end{eqnarray}
We note that in \textit{DT} we can explicitly discuss what happens when the
fields are at equal times, whereas in \textit{CT} field theory, we tend to
avoid this in discussions of time-ordered products, appealing to the
properties of distributions.

Now considering the quantum operator equation of motion 
\begin{equation}
\hat{\omega}^{-1}\hat{\psi}_{n+1}+\frac{4\kappa }{|\omega |}\hat{\psi}_n+%
\hat{\omega}\hat{\psi}_{n-1}=\frac{6T}{|\omega |}\eta _n,
\end{equation}
we take its matrix element between the ground state, use the above relations 
$\left( \ref{aa}-\ref{bb}\right) $ and hence find the ground state
functional 
\begin{equation}
Z\left[ \eta \right] \equiv \langle 0|0\rangle _\eta =\exp \left\{
-iT^2\sum_{r,s=-\infty }^\infty \bar{\eta}_rS_F^{r-s}\eta _s\right\} ,
\end{equation}
suppressing Dirac space indices. Then using the result 
\begin{equation}
\langle 0|\tilde{T}\bar{\psi}_{na}\hat{\psi}_{mb}|0\rangle =\lim_{\eta
\rightarrow 0}\left\{ \frac iT\frac \partial {\partial \eta _{na}}\frac{-i}%
T\frac \partial {\partial \bar{\eta}_{mb}}Z\left[ \eta \right] \right\}
\end{equation}
explicitly showing the Dirac space indices, we find 
\begin{equation}
\langle 0|\tilde{T}\bar{\psi}_{na}\hat{\psi}_{mb}|0\rangle =-iS_{F\;ba}^{m-n}
\end{equation}
or 
\begin{equation}
\langle 0|\tilde{T}\hat{\psi}_{na}^{+}\hat{\psi}_{mb}|0\rangle
=-i(S_{F\;}^{m-n}\gamma ^0)_{ba}.
\end{equation}
Using our results for the \textit{DT} Dirac propagator in the previous
subsection and choosing appropriate temporal indices $m$ and $n$, we extract
the following information: 
\begin{eqnarray}
n>m:\langle 0|\psi _n^{+}\psi _m|0\rangle &=&-\frac 3{\sqrt{9-3\kappa ^2}%
}e^{-i\left( n-m\right) \theta }\hat{\omega}^{m-n}\gamma ^0  \nonumber \\
&=&\frac 3{\sqrt{9-3\kappa ^2}}\left[ 
\begin{array}{cc}
-e^{-i\left( n-m\right) \sigma } & 0 \\ 
0 & e^{-i\left( n-m\right) \delta }
\end{array}
\right]  \label{L1} \\
\langle 0|\psi _m^{+}\psi _n|0\rangle &=&-\frac 3{\sqrt{9-3\kappa ^2}}\gamma
^0e^{i\left( n-m\right) \theta }\hat{\omega}^{n-m}  \nonumber \\
&=&\frac 3{\sqrt{9-3\kappa ^2}}\left[ 
\begin{array}{cc}
-e^{i\left( n-m\right) \sigma } & 0 \\ 
0 & e^{i\left( n-m\right) \delta }
\end{array}
\right] \\
n<m:\langle 0|\psi _m\psi _n^{+}|0\rangle &=&\frac 3{\sqrt{9-3\kappa ^2}%
}e^{-i\left( m-n\right) \theta }\hat{\omega}^{m-n}\gamma ^0  \nonumber \\
&=&\frac 3{\sqrt{9-3\kappa ^2}}\left[ 
\begin{array}{cc}
e^{i\left( n-m\right) \delta } & 0 \\ 
0 & -e^{i\left( n-m\right) \sigma }
\end{array}
\right] \\
\langle 0|\psi _n^{+}\psi _m|0\rangle &=&\frac 3{\sqrt{9-3\kappa ^2}%
}e^{i\left( m-n\right) \theta }\gamma ^0\hat{\omega}^{n-m}  \nonumber \\
&=&\frac 3{\sqrt{9-3\kappa ^2}}\left[ 
\begin{array}{cc}
e^{-i\left( n-m\right) \delta } & 0 \\ 
0 & -e^{-i\left( n-m\right) \sigma }
\end{array}
\right] .  \label{L2}
\end{eqnarray}

In addition, by taking $n=m$ we find 
\begin{equation}
\langle 0|\left[ \hat{\psi}_n,\hat{\psi}_n^{+}\right] |0\rangle =\frac{%
3\gamma ^0}{2\sqrt{9-3\kappa ^2}}.
\end{equation}
By using the \textit{DT} operator equation of motion for the source free
case 
\begin{equation}
\hat{\omega}^{-1}\hat{\psi}_{n+1}+\frac{4\kappa }{|\omega |}\hat{\psi}_n+%
\hat{\omega}\hat{\psi}_{n-1}=0
\end{equation}
we deduce 
\begin{equation}
\langle 0|\hat{\psi}_n\hat{\psi}_n^{+}|0\rangle =-\langle 0|\hat{\psi}_n^{+}%
\hat{\psi}_n|0\rangle =\frac{3\gamma ^0}{\sqrt{9-3\kappa ^2}}.
\end{equation}
We now use the above ground state expectation values and the reasonable
assumption that free field anticommutators are c-numbers to write down the
following anticommutators: 
\begin{eqnarray}
\left\{ \hat{\psi}_{n+1}^{+},\hat{\psi}_n\right\} &=&\frac{6i\gamma ^0}{%
|\omega |}\hat{\omega}^{-1},  \label{a1} \\
\left\{ \hat{\psi}_n^{+},\hat{\psi}_{n+1}\right\} &=&\frac{-6i\gamma ^0}{%
|\omega |}\hat{\omega},  \label{a2} \\
\left\{ \hat{\psi}_n^{+},\hat{\psi}_n\right\} &=&0,  \label{surprise}
\end{eqnarray}
knowing they are consistent with the operator equations of motion and with
the propagator. This amounts to our quantisation prescription.

We note that the last result, (\ref{surprise}) is surprising considering
that the canonical \textit{CT} equal-time anticommutator of Dirac field
operators is non-zero, i.e. 
\begin{equation}
\left\{ \hat{\psi}\left( t\right) _a,\hat{\psi}^{+}\left( t\right)
_b\right\} =\delta _{ab}.  \label{can}
\end{equation}
There are two points to be made here:

\begin{enumerate}
\item  The Schwinger method in \textit{CT} field theory does not lead to
equal time anticommutators directly. It deals with time-ordered products,
which involve the Heaviside distribution (the step function) and then we
have to consider temporal limits carefully. So for example the \textit{CT }%
analogues of the results $\left( \ref{L1}-\ref{L2}\right) \;$are 
\begin{eqnarray}
t>0:\;\;\;\langle 0|\hat{\psi}_a^{+}\left( t\right) \hat{\psi}_b\left(
0\right) |0\rangle =\left[ 
\begin{array}{cc}
0 & 0 \\ 
0 & e^{-imt}
\end{array}
\right] _{ba}  \nonumber \\
\langle 0|\hat{\psi}_b^{+}\left( 0\right) \hat{\psi}_a\left( t\right)
|0\rangle =\left[ 
\begin{array}{cc}
0 & 0 \\ 
0 & e^{imt}
\end{array}
\right] _{ba}  \nonumber \\
t<0:\;\;\;\langle 0|\hat{\psi}_b\left( 0\right) \hat{\psi}_a^{+}\left(
t\right) |0\rangle =\left[ 
\begin{array}{cc}
e^{imt} & 0 \\ 
0 & 0
\end{array}
\right] _{ba} \\
\langle 0|\hat{\psi}_a\left( t\right) \hat{\psi}_b^{+}\left( 0\right)
|0\rangle =\left[ 
\begin{array}{cc}
e^{-imt} & 0 \\ 
0 & 0
\end{array}
\right] _{ba},  \nonumber
\end{eqnarray}
from which we deduce 
\begin{equation}
\lim_{t\rightarrow 0^{+}}\left\{ \langle 0|\hat{\psi}_a^{+}\left( t\right) 
\hat{\psi}_b\left( 0\right) |0\rangle +\langle 0|\hat{\psi}_b\left( 0\right) 
\hat{\psi}_a^{+}\left( t\right) |0\rangle \right\} =\delta _{ab,}^{}
\end{equation}
which is consistent with $\left( \ref{can}\right) ;$

\item  The value zero in $\left( \ref{surprise}\right) $ is explained by the
presence of oscillon solutions, which also manifest themselves in the
relations $\left( \ref{L1}-\ref{L2}\right) .$ This suggests that the \textit{%
DT} field operator $\hat{\psi}_n$ is more complicated than its \textit{CT}
analogue $\hat{\psi}\left( t\right) .$
\end{enumerate}

\subsection{The free particle Dirac equation}

We turn now to solutions to the free particle \textit{DT} Dirac equation 
\begin{equation}
\hat{\omega}^{-1}\psi _{n+1}+\frac{4\kappa }{|\omega |}\psi _n+\hat{\omega}%
\psi _n\stackunder{c}{=}0.
\end{equation}
If we write$\;\psi _n=\hat{\omega}^n\phi _n$ then 
\begin{equation}
\left( U_n-2\eta +U_n^{-1}\right) \phi _n\stackunder{c}{=}0,\;\;\;\;\eta =%
\frac{-2\kappa }{|\omega |}.
\end{equation}
Assuming this equation has solutions of the form$\;\phi _n=z^n\chi ,$ where $%
\chi $ is a two component object and $z$ is taken complex and non-zero, then
we find 
\begin{equation}
\left( z-2\eta +z^{-1}\right) z^n\chi \stackunder{c}{=}0,  \label{SD}
\end{equation}
from which we deduce$\;z=e^{\pm i\theta }$,\ where $\cos \theta =\eta $
satisfies $\left( \ref{theta}\right) .$ We will assume we are in the
elliptic regime, so that $\theta $ is real. Next, define the two component
Dirac space basis functions 
\begin{equation}
u\equiv \left[ 
\begin{array}{c}
\sqrt{2m} \\ 
0
\end{array}
\right] ,\;\;\;\;\;\,v\equiv \left[ 
\begin{array}{c}
0 \\ 
\sqrt{2m}
\end{array}
\right] .
\end{equation}
Then solutions to (\ref{SD}) may be written in the form 
\begin{equation}
\phi _n=\frac 1{2m}\left\{ [ae^{-in\theta }+c^{*}e^{in\theta
}]u+[de^{-in\theta }+b^{*}e^{in\theta }]v\right\} .
\end{equation}
We note that there are four independent solutions, as discussed above. Hence
finally we get 
\begin{eqnarray}
\psi _n &=&\hat{\omega}^n\phi _n  \nonumber \\
&=&\frac 1{2m}\left\{ [ae^{-in\delta }+c^{*}e^{in\sigma }]u+[de^{-in\sigma
}+b^{*}e^{in\delta }]v\right\} .
\end{eqnarray}
Second quantisation turns the coefficients $a,b,c,d,$ etc. into operators,
so that the solutions to the quantised \textit{DT} Dirac equation are 
\begin{eqnarray}
\hat{\psi}_n &=&\frac 1{2m}\left\{ [\hat{a}e^{-in\delta }+\hat{c}%
^{+}e^{in\sigma }]u+[\hat{d}e^{-in\sigma }+\hat{b}^{+}e^{in\delta
}]v\right\} ,  \nonumber \\
\hat{\psi}_n^{+} &=&\frac 1{2m}\left\{ u^{+}[\hat{a}^{+}e^{in\delta }+\hat{c}%
e^{-in\sigma }]+v^{+}[\hat{d}^{+}e^{in\sigma }+\hat{b}e^{-in\delta
}]\right\} .
\end{eqnarray}
We may invert the relationship (assuming we are in the elliptic regime) to
find particle-antiparticle creation and annihilation operators: 
\begin{eqnarray}
\hat{a} &=&\frac i{2\sin \theta }u^{+}\left\{ \hat{\psi}_{n+1}e^{-i\xi }-%
\hat{\psi}_ne^{i\theta }\right\} e^{in\delta },\;\;\;\;\;\stackunder{T}{%
\rightarrow }u^{+}\hat{\psi}\left( t\right) e^{imt}  \nonumber \\
\hat{a}^{+} &=&\frac{-i}{2\sin \theta }\left\{ \hat{\psi}_{n+1}^{+}e^{i\xi }-%
\hat{\psi}_n^{+}e^{-i\theta }\right\} ue^{-in\delta },\;\;\;\;\;\stackunder{T%
}{\rightarrow }\hat{\psi}^{+}\left( t\right) ue^{-imt} \\
\hat{b} &=&\frac i{2\sin \theta }\left\{ \hat{\psi}_{n+1}^{+}e^{-i\xi }-\hat{%
\psi}_n^{+}e^{i\theta }\right\} ve^{in\delta },\;\;\;\,\;\;\;\stackunder{T}{%
\rightarrow }\hat{\psi}^{+}\left( t\right) ve^{imt}  \nonumber \\
\hat{b}^{+} &=&\frac{-i}{2\sin \theta }v^{+}\left\{ \hat{\psi}_{n+1}e^{i\xi
}-\hat{\psi}_ne^{-i\theta }\right\} e^{-in\delta },\;\;\,\;\stackunder{T}{%
\rightarrow }v^{+}\hat{\psi}\left( t\right) e^{-imt}
\end{eqnarray}
and oscillon/anti-oscillon ladder operators: 
\begin{eqnarray}
\hat{c} &=&\frac i{2\sin \theta }\left\{ \hat{\psi}_{n+1}^{+}e^{i\xi }-\hat{%
\psi}_n^{+}e^{i\theta }\right\} ue^{in\sigma },\;\;\;\;\;\stackunder{T}{%
\rightarrow }0  \nonumber \\
\hat{c}^{+} &=&\frac{-i}{2\sin \theta }u^{+}\left\{ \hat{\psi}_{n+1}e^{-i\xi
}-\hat{\psi}_ne^{-i\theta }\right\} e^{-in\sigma },\;\stackunder{T}{%
\rightarrow 0} \\
\hat{d} &=&\frac i{2\sin \theta }v^{+}\left\{ \hat{\psi}_{n+1}e^{i\xi }-\hat{%
\psi}_ne^{i\theta }\right\} e^{in\sigma },\;\;\;\;\stackunder{T}{\rightarrow 
}0  \nonumber \\
\hat{d}^{+} &=&\frac{-i}{2\sin \theta }\left\{ \hat{\psi}_{n+1}^{+}e^{-i\xi
}-\hat{\psi}_n^{+}e^{-i\theta }\right\} ve^{-in\sigma },\;\;\stackunder{T}{%
\rightarrow }0.
\end{eqnarray}
In the above equations we show what should happen in the limit 
\begin{equation}
\stackunder{T}{\rightarrow }\;\equiv \lim \{T\rightarrow 0,\;n\rightarrow
\infty ,\;n\theta \rightarrow t\}
\end{equation}
assuming for example 
\begin{equation}
\psi _{n+1}\simeq \psi \left( t\right) +T\dot{\psi}\left( t\right) +O\left(
T^2\right) ,\;\;\;\text{etc}
\end{equation}
We note that we have not indexed the ladder operators with a temporal index $%
n$, but it is implied. These operators have the same temporal properties as
the creation and annihilation operators found for the \textit{DT} bosonic
oscillator discussed in \textit{Principles I}. Although such operators
satisfy relations such as $\hat{a}_n=\hat{a}_{n+1}$ for example, they are
not actually invariants of the motion in that they do not commute with the
timestep operator $\hat{U}_n$. In the terminology of \textit{Principles I},
these operators are not compatible with the timestep operator. The
resolution of this apparent paradox is to note that we are actually working
in the Heisenberg picture.

To understand further the basic properties of these operators, we consider
the sixty four possible ground state expectation values of products of the
eight operators $\hat{a},\;\hat{a}^{+},...,\hat{d}^{+}.\;$Of these, only
four are non-zero$.$ We find 
\begin{eqnarray}
\langle 0|\hat{a}\hat{a}^{+}|0\rangle &=&\langle 0|\hat{b}\hat{b}%
^{+}|0\rangle =\frac{6m}{\sqrt{9-3\kappa ^2}},  \nonumber \\
\langle 0|\hat{c}\hat{c}^{+}|0\rangle &=&\langle 0|\hat{d}\hat{d}%
^{+}|0\rangle =\frac{-6m}{\sqrt{9-3\kappa ^2}}.
\end{eqnarray}
This is consistent with the view that the operators $\hat{a},\;\hat{b},\;%
\hat{c},$ and $\hat{d}$ annihilate the ground state ket $|0\rangle $ and
that their adjoints annihilate the ground state bra $\langle 0|$.

If now we use the anticommutation relations $\left( \ref{a1}-\ref{surprise}%
\right) $ for the fields worked out previously, we find the following
non-zero anticommutators: 
\begin{eqnarray}
\left\{ \hat{a},\hat{a}^{+}\right\} &=&\left\{ \hat{b},\hat{b}^{+}\right\} =%
\frac{6m}{\sqrt{9-3\kappa ^2}},\;\;\;\stackunder{T}{\rightarrow }2m, \\
\left\{ \hat{c},\hat{c}^{+}\right\} &=&\left\{ \hat{d},\hat{d}^{+}\right\} =%
\frac{-6m}{\sqrt{9-3\kappa ^2}},\;\;\;\;\;\stackunder{T}{\rightarrow }-2m.
\end{eqnarray}
The fundamental result here is that the particle and anti-particle creation
and annihilation operators lead to states with a positive inner product,
whereas the oscillon and anti-oscillon operators lead to negative norm
states, which means that oscillons and anti-oscillons should be regarded as
unphysical.

Finally, we find that the conserved charge $\hat{Q}$ operator$,$ given by 
\begin{eqnarray}
\hat{Q} &=&\frac 1{12}\hat{\psi}_{n+1}^{+}\left[ 3-i\kappa \gamma ^0\right] 
\hat{\psi}_n+\frac 1{12}\hat{\psi}_n^{+}\left[ 3+i\kappa \gamma ^0\right] 
\hat{\psi}_{n+1}  \nonumber \\
&&-\frac 1{12}\hat{\psi}_n\left[ 3-i\kappa \gamma ^0\right] \hat{\psi}%
_{n+1}^{+}-\frac 1{12}\hat{\psi}_{n+1}\left[ 3+i\kappa \gamma ^0\right] \hat{%
\psi}_n^{+}
\end{eqnarray}
may be written in the form 
\begin{equation}
\hat{Q}=\frac{\sqrt{9-3\kappa ^2}}{6m}\left[ \hat{a}^{+}\hat{a}-\hat{b}^{+}%
\hat{b}+\hat{c}^{+}\hat{c}-\hat{d}^{+}\hat{d}\right] .
\end{equation}
Taking the anticommutators into account, we see that particles,
antiparticles, oscillons and anti-oscillons carry the same magnitude of
charge. We expect therefore that whilst oscillons and anti-oscillons would
not appear as external \textit{in} or \textit{out} asymptotic particles,
they can carry charge, linear momentum and spin, and should contribute in
Feynman diagrams. In particular, there may be significant novel effects,
with particles making virtual transitions to oscillons and back again. When
we come to the field theory, discussed next, we expect to find that electric
charge as a physical observable should only be associated with particles or
antiparticles with momenta below the parabolic barrier.

\section{The Dirac equation in 1+3 dimensions}

We turn now to the Dirac equation in one time and three spatial dimensions.
First we consider the free particle system. Using the virtual paths
discussed previously \textit{\ }the \textit{CT }Dirac Lagrangian density 
\begin{equation}
\mathcal{L}\left( x\right) =\frac{_1}{^2}i\bar{\psi}\left( x\right) 
\overrightarrow{\partial \!\!\!\!\!\!\diagup }\psi \left( x\right) -\frac{_1%
}{^2}i\bar{\psi}\left( x\right) \overleftarrow{\partial \!\!\!\!\!\!\diagup }%
\psi \left( x\right) -m\bar{\psi}\left( x\right) \psi \left( x\right)
\end{equation}
gives the system function density 
\begin{eqnarray}
\mathcal{F}^n\left( \mathbf{x}\right) &=&\frac{_1}{^2}i\left\{ \psi
_n^{+}\left( \mathbf{x}\right) \psi _{n+1}\left( \mathbf{x}\right) -\psi
_{n+1}^{+}\left( \mathbf{x}\right) \psi _n\left( \mathbf{x}\right) \right\} 
\nonumber \\
&&-\frac T6\left\{ 
\begin{array}{c}
2\psi _{n+1}^{+}\left( \mathbf{x}\right) \overleftrightarrow{H}\psi
_{n+1}\left( \mathbf{x}\right) +\psi _n^{+}\left( \mathbf{x}\right) 
\overleftrightarrow{H}\psi _{n+1}\left( \mathbf{x}\right) \\ 
+\psi _{n+1}^{+}\left( \mathbf{x}\right) \overleftrightarrow{H}\psi _n\left( 
\mathbf{x}\right) +2\psi _n^{+}\left( \mathbf{x}\right) \overleftrightarrow{H%
}\psi _n\left( \mathbf{x}\right)
\end{array}
\right\} ,  \label{fden}
\end{eqnarray}
where 
\begin{equation}
\overleftrightarrow{H}\equiv \frac{_1}{^2}i\overleftarrow{\nabla }\mathbf{%
\cdot \alpha }-\frac{_1}{^2}i\mathbf{\alpha \cdot }\overrightarrow{\nabla }%
+m\beta
\end{equation}
is the standard Dirac one-particle Hamiltonian operator. The Cadzow equation
for $\psi _n\left( \mathbf{x}\right) $ is formally given by 
\begin{equation}
\frac \partial {\partial \psi _n^{+}}\left\{ \mathcal{F}^n+\mathcal{F}%
^{n-1}\right\} \stackunder{c}{=}\nabla \cdot \frac \partial {\partial \nabla
\psi _n^{+}}\left\{ \mathcal{F}^n+\mathcal{F}^{n-1}\right\} ,
\end{equation}
which gives the Cadzow equation for $\psi _n$: 
\begin{equation}
i\frac{\psi _{n+1}-\psi _{n-1}}{2T}\stackunder{c}{=}\frac{_1}{^6}%
\overrightarrow{H}\left\{ \psi _{n+1}+4\psi _n+\psi _{n-1}\right\} ,
\end{equation}
where $\overrightarrow{H}\equiv -i\mathbf{\alpha \cdot }\overrightarrow{%
\nabla }+m\beta .$ In this form we can readily identify the various parts of
the non-covariant form of the \textit{CT} Dirac equation. Another form is 
\begin{equation}
\overrightarrow{\Lambda ^{+}}\psi _{n+1}\stackunder{c}{=}-\frac{4iT}6%
\overrightarrow{H}\psi _n+\overrightarrow{\Lambda ^{-}}\psi _{n-1}
\label{xeqm}
\end{equation}
where $\overrightarrow{\Lambda ^{+}}\equiv \frac{_1}{^2}+\frac{iT}6%
\overrightarrow{H}$, $\overrightarrow{\Lambda ^{-}}\equiv \frac{_1}{^2}-%
\frac{iT}6\overrightarrow{H}.$ Likewise, we find the Cadzow equation for $%
\psi _n^{+}$

\begin{equation}
\psi _{n+1}^{+}\overleftarrow{\Lambda ^{-}}\stackunder{c}{=}\frac{4iT}6\psi
_n^{+}\overleftarrow{H}+\psi _{n-1}^{+}\overleftarrow{\Lambda ^{+}},
\label{yeqm}
\end{equation}
where$\;\overleftarrow{H}\equiv i\overleftarrow{\nabla }\mathbf{\cdot \alpha 
}+m\beta $,$\;\overleftarrow{\Lambda ^{+}}\equiv \frac 12+\frac{iT}6%
\overleftarrow{H},\;$and$\;\overleftarrow{\Lambda ^{-}}\equiv \frac 12-\frac{%
iT}6\overleftarrow{H}.$ The equations in the form involving the $\Lambda $
operators are useful for proving that various invariants of the motion are
indeed constant in time.

With the Fourier transforms 
\begin{equation}
\tilde{\psi}_n\left( \mathbf{p}\right) \equiv \int d^3\mathbf{x}e^{-i\mathbf{%
p\cdot x}}\psi _n^{}\left( \mathbf{x}\right) ,\;\;\;\tilde{\psi}_n^{+}\left( 
\mathbf{p}\right) \equiv \int d^3\mathbf{x}e^{i\mathbf{p\cdot x}}\psi
_n^{+}\left( \mathbf{x}\right) ,  \label{Fourier}
\end{equation}
we find 
\begin{eqnarray}
&&\tilde{\Lambda}^{+}\left( \mathbf{p}\right) \tilde{\psi}_{n+1}\left( 
\mathbf{p}\right) \stackunder{c}{=}\frac{-4i}6TH\left( \mathbf{p}\right) 
\tilde{\psi}_n\left( \mathbf{p}\right) +\tilde{\Lambda}^{-}\left( \mathbf{p}%
\right) \tilde{\psi}_{n-1}\left( \mathbf{p}\right) ,  \nonumber \\
&&\tilde{\psi}_{n+1}^{+}\left( \mathbf{p}\right) \tilde{\Lambda}^{-}\left( 
\mathbf{p}\right) \stackunder{c}{=}\frac{4i}6T\tilde{\psi}_n^{+}\left( 
\mathbf{p}\right) H\left( \mathbf{p}\right) +\tilde{\psi}_{n-1}^{+}\left( 
\mathbf{p}\right) \tilde{\Lambda}^{+}\left( \mathbf{p}\right)  \label{eqmp}
\end{eqnarray}
where $H\left( \mathbf{p}\right) \equiv \mathbf{\alpha \cdot p}+m\beta ,$ $%
\tilde{\Lambda}^{+}\left( \mathbf{p}\right) \equiv \frac{_1}{^2}+\frac{iT}%
6H\left( \mathbf{p}\right) ,$ and $\tilde{\Lambda}^{-}\left( \mathbf{p}%
\right) \equiv \frac{_1}{^2}-\frac{iT}6H\left( \mathbf{p}\right) .$ Now
consider the construction 
\begin{equation}
C^n\equiv \int \frac{d^3\mathbf{p}}{\left( 2\pi \right) ^3}\mathcal{C}\left( 
\mathbf{p}\right) \left\{ \tilde{\psi}_n\left( \mathbf{p}\right) \tilde{%
\Lambda}^{+}\left( \mathbf{p}\right) \tilde{\psi}_{n+1}\left( \mathbf{p}%
\right) +\tilde{\psi}_{n+1}\left( \mathbf{p}\right) \tilde{\Lambda}%
^{-}\left( \mathbf{p}\right) \tilde{\psi}_n\left( \mathbf{p}\right) \right\}
,  \label{Logan}
\end{equation}
where $\mathcal{C}\left( \mathbf{p}\right) $ is arbitrary. Using the
equations of motion $\left( \ref{eqmp}\right) $ we readily find 
\begin{equation}
C^n\stackunder{c}{=}C^{n-1},
\end{equation}
which shows that $C^n$ is a Logan invariant \cite{J&N-I,LOGAN.73} of the
system.

\subsection{The \textit{DT }Dirac propagator}

In the presence of external fermionic sources the system function density
becomes 
\begin{equation}
\mathcal{F}_\eta ^n\left( \mathbf{x}\right) =\mathcal{F}_{}^n\left( \mathbf{x%
}\right) +\frac{_1}{^2}T\left\{ 
\begin{array}{c}
\bar{\eta}_n\left( \mathbf{x}\right) \psi _n\left( \mathbf{x}\right) +\bar{%
\eta}_{n+1}\left( \mathbf{x}\right) \psi _{n+1}\left( \mathbf{x}\right) \\ 
+\bar{\psi}_n\left( \mathbf{x}\right) \eta _n\left( \mathbf{x}\right) +\bar{%
\psi}_{n+1}\left( \mathbf{x}\right) \eta _{n+1}\left( \mathbf{x}\right)
\end{array}
\right\} ,
\end{equation}
which gives the Cadzow equation for $\psi _n$: 
\begin{equation}
i\frac{\psi _{n+1}-\psi _{n-1}}{2T}\stackunder{c}{=}\frac{_1}{^6}%
\overrightarrow{H}\left\{ \psi _{n+1}+4\psi _n+\psi _{n-1}\right\} -\beta
\eta _n.  \label{ceq}
\end{equation}
Likewise, we find the Cadzow equation for $\psi _n^{+}:$

\begin{equation}
\psi _{n+1}^{+}\overleftarrow{\Lambda ^{-}}\stackunder{c}{=}\frac{4iT}6\psi
_n^{+}\overleftarrow{H}+\psi _{n-1}^{+}\overleftarrow{\Lambda ^{+}}%
\stackunder{}{-iT}\eta _n^{+}\beta .  \label{eqm2}
\end{equation}
Taking Fourier transforms and with the definitions 
\begin{equation}
\hat{\Omega}\left( \mathbf{p}\right) \equiv \frac{\kappa +3i\hat{H}\left( 
\mathbf{p}\right) }{\sqrt{9+\kappa ^2}},\;\;\;\;\;\kappa \equiv TE_{\mathbf{p%
}},\;\;\hat{H}\left( \mathbf{p}\right) \equiv \frac{H\left( \mathbf{p}%
\right) }{E_{\mathbf{p}}}
\end{equation}
the equation $\left( \ref{ceq}\right) $ becomes 
\begin{equation}
\left( \hat{\Omega}^{+}\left( \mathbf{p}\right) U_n-2\eta _E+\hat{\Omega}%
\left( \mathbf{p}\right) U_n^{-1}\right) \tilde{\psi}_n\left( \mathbf{p}%
\right) \stackunder{c}{=}T\Gamma _E\hat{H}\left( \mathbf{p}\right) \beta 
\tilde{\eta}_n\left( \mathbf{p}\right) ,\;\;
\end{equation}
where 
\begin{equation}
\Gamma _E\equiv \frac 6{\sqrt{9+\kappa ^2}},\;\;\;\;\;\eta _E=\frac{-2\kappa 
}{\sqrt{9+\kappa ^2}}.
\end{equation}
With our experience of the \textit{DT} Dirac equation in $1+0$ dimensions,
we may immediately write down the formal solution 
\begin{equation}
\tilde{\psi}_n\left( \mathbf{p}\right) =\tilde{\psi}_n^{\left( 0\right)
}\left( \mathbf{p}\right) -T\sum_{m=-\infty }^\infty \tilde{S}_F^{n-m}\left( 
\mathbf{p}\right) \tilde{\eta}_m\left( \mathbf{p}\right) ,
\end{equation}
where $\tilde{\psi}_n^{\left( 0\right) }\left( \mathbf{p}\right) $ is a
solution to the free \textit{DT} Dirac equation and the propagators 
\begin{equation}
\tilde{S}_F^n\left( \mathbf{p}\right) \equiv \tilde{\Delta}_F^n\left( 
\mathbf{p}\right) \hat{\Omega}^n\left( \mathbf{p}\right) \hat{H}\left( 
\mathbf{p}\right) \gamma ^0,\;\;\;\;\;\tilde{\Delta}_F^n\left( \mathbf{p}%
\right) \equiv \frac{e^{-i|n|\theta _E}\Gamma _E}{2i\sin \theta _E}
\label{star}
\end{equation}
satisfy the equations 
\begin{eqnarray}
\left( \hat{\Omega}^{-1}\left( \mathbf{p}\right) U_n-2\eta _E+\hat{\Omega}%
\left( \mathbf{p}\right) U_n^{-1}\right) \tilde{S}_F^n\left( \mathbf{p}%
\right) &=&-\Gamma _E\hat{H}\left( \mathbf{p}\right) \beta \delta _n, \\
\left( U_n-2\eta _E+U_n^{-1}\right) \tilde{\Delta}_F^n\left( \mathbf{p}%
\right) &=&-\Gamma _E\delta _n.
\end{eqnarray}
Note that the bosonic propagator $\tilde{\Delta}_F^n\left( \mathbf{p}\right) 
$ in $\left( \ref{star}\right) $ is dimensionless whereas the bosonic
propagator in $\left( \ref{sta}\right) $ has the physical dimensions of a
length in our system of units.

To investigate the nature of the propagator, define the Fourier series
transform 
\begin{equation}
\tilde{S}_F\left( \mathbf{p},\Theta \right) \equiv T\sum_{n=-\infty }^\infty
e^{in\Theta }\tilde{S}_F^n\left( \mathbf{p}\right) ,
\end{equation}
where the parameter $\Theta $ is taken real. Then we find 
\begin{equation}
\left[ \kappa \left( \cos \Theta +2\right) \hat{H}-3\sin \Theta \right] 
\tilde{S}_F\left( \mathbf{p},\Theta \right) =-3T\beta .
\end{equation}
We may now solve for $\tilde{S}_F\left( \mathbf{p},\Theta \right) $ if we
give $\kappa ^2$ a small imaginary term, according to the standard Feynman $%
m\rightarrow m-i\epsilon $ prescription. Hence we find 
\begin{equation}
\tilde{S}_F\left( \mathbf{p},\Theta \right) =\frac{-3T\left[ \kappa \left(
\cos \Theta +2\right) \hat{H}-3\sin \Theta \right] \beta }{\left[ \kappa
^2(\cos \Theta +2)^2-9\sin ^2\Theta -i\epsilon \right] }.
\end{equation}
In this form the propagator looks quite different to the standard \textit{CT}
propagator, but a suitable reparametrisation can change this. We introduce
the parameter $p_0$ (which should \textbf{not} be confused with $E_p\equiv 
\sqrt{\mathbf{p\cdot p}+m^2})$ related to the parameter $\Theta $ by 
\begin{equation}
\cos \Theta =\frac{6-2p_0^2T^2}{6+p_0^2T^2},\;\;\;sign(\Theta )=sign\left(
p_0\right) .  \label{param}
\end{equation}
Then we find 
\begin{equation}
\tilde{S}_F\left( \mathbf{p},\Theta \right) =\frac{\gamma ^0p_0+\gamma
^ip_i+m}{p_0^2-E_p^2+i\epsilon }+O\left( T^2\right) =\frac{\not{p}+m}{%
p^2-m^2+i\epsilon }+O\left( T^2\right) .
\end{equation}
From this we see that our propagator indeed looks exactly like the \textit{CT%
} Feynman propagator for the Dirac field in lowest order in $T$.

This result is important for two reasons. First, the reparametrisation $%
\left( \ref{param}\right) $ is precisely the same as the parametrisation $%
\left( \ref{bparam}\right) $ used in the analogous expansion for the bosonic
propagator, (\ref{bprop})$.$ This means that the parameter $\Theta $ flowing
through Feynman diagram networks has the same representation and
interpretation for fermions as it has for bosons, and can be justifiably
regarded as the \textit{DT} analogue of energy, up to a factor of $T$. In 
\textit{Principles III} we found that for scalar field theory, the sum of
incoming $\Theta $ parameters was conserved at each vertex in a Feynman
diagram, and so we conjecture that an analogous result holds for higher spin
fields. The second important point is that this results shows us that
Lorentz covariance for the Dirac equation emerges from \textit{DT} mechanics
at the same level of approximation as it does in bosonic theory. There is
every reason, therefore to regard the loss of manifest Lorentz covariance in
the theory as not a serious problem, provided that we work in the regime
where $T$ is close to zero in an appropriate sense.

\subsection{Field anticommutators}

With the Fourier series transforms implied by $\left( \ref{Fourier}\right) $
the action sum in the presence of external sources becomes 
\begin{eqnarray}
\hat{A}^{NM}\left[ \eta \right] &\equiv &\sum_{n=M}^{N-1}\int d\mathbf{x}%
\mathcal{F}^n\left[ \eta \right]  \nonumber \\
&=&\hat{A}^{NM}+T\sum_{n=M+1}^{N-1}\int \frac{d\mathbf{p}}{\left( 2\pi
\right) ^3}\left[ \tilde{\eta}_n^{+}\left( \mathbf{p}\right) \gamma ^0\tilde{%
\psi}_n\left( \mathbf{p}\right) +\tilde{\psi}_n^{+}\left( \mathbf{p}\right)
\gamma ^0\tilde{\eta}_n\left( \mathbf{p}\right) \right]  \nonumber \\
&&+\frac{_1}{^2}T\int \frac{d\mathbf{p}}{\left( 2\pi \right) ^3}\{\tilde{\eta%
}_N^{+}\left( \mathbf{p}\right) \gamma ^0\tilde{\psi}_N\left( \mathbf{p}%
\right) +\tilde{\psi}_N^{+}\left( \mathbf{p}\right) \gamma ^0\tilde{\eta}%
_N\left( \mathbf{p}\right)  \nonumber \\
&&\;\;\;\;\;\;\;+\tilde{\eta}_M^{+}\left( \mathbf{p}\right) \gamma ^0\tilde{%
\psi}_M\left( \mathbf{p}\right) +\tilde{\psi}_M^{+}\left( \mathbf{p}\right)
\gamma ^0\tilde{\eta}_M\left( \mathbf{p}\right) \}.
\end{eqnarray}
For the Schwinger action principle, we first define the functional
derivatives 
\begin{equation}
\frac \delta {\delta \eta _n\left( \mathbf{x}\right) }\eta _m\left( \mathbf{y%
}\right) =\delta _{n-m}\delta ^3\left( \mathbf{x-y}\right) ,\;\;\;\frac
\delta {\delta \eta _n^{+}\left( \mathbf{x}\right) }\eta _m^{+}\left( 
\mathbf{y}\right) =\delta _{n-m}\delta ^3\left( \mathbf{x-y}\right)
\end{equation}
and 
\begin{equation}
\frac \delta {\delta \tilde{\eta}_n^{+}\left( \mathbf{p}\right) }\equiv \int
d\mathbf{x}e^{-i\mathbf{p\cdot x}}\frac \delta {\delta \eta _n^{+}\left( 
\mathbf{x}\right) },\;\;\;\;\;\frac \delta {\delta \tilde{\eta}_n\left( 
\mathbf{p}\right) }\equiv \int d\mathbf{x}e^{i\mathbf{p\cdot x}}\frac \delta
{\delta \eta _n\left( \mathbf{x}\right) }
\end{equation}
so that 
\begin{eqnarray}
\frac \delta {\delta \tilde{\eta}_n\left( \mathbf{p}\right) }\tilde{\eta}%
_m\left( \mathbf{q}\right) &=&\left( 2\pi \right) ^3\delta ^3\left( \mathbf{%
p-q}\right) \delta _{n-m},  \nonumber \\
\frac \delta {\delta \tilde{\eta}_n^{+}\left( \mathbf{p}\right) }\tilde{\eta}%
_m^{+}\left( \mathbf{q}\right) &=&\left( 2\pi \right) ^3\delta ^3\left( 
\mathbf{p-q}\right) \delta _{n-m}.
\end{eqnarray}
Then the Schwinger action principle gives 
\begin{eqnarray}
\frac{-i}T\frac \delta {\delta \tilde{\eta}_n^{+}\left( \mathbf{p}\right)
}\langle \alpha ,N|\beta ,M\rangle _\eta &=&\langle \alpha ,N|\gamma ^0%
\tilde{\psi}_n\left( \mathbf{p}\right) |\beta ,M\rangle _\eta ,\;\;\;N>n>M 
\nonumber \\
\frac iT\frac \delta {\delta \tilde{\eta}_n^{}\left( \mathbf{p}\right)
}\langle \alpha ,N|\beta ,M\rangle _\eta &=&\langle \alpha ,N|\stackrel{}{%
\tilde{\psi}_n^{+}\left( \mathbf{p}\right) \gamma ^0}|\beta ,M\rangle _\eta
,\;\;\;N>n>M
\end{eqnarray}
and so on. With this and Cadzow's equations of motion we find the vacuum
functional 
\begin{equation}
Z\left[ \eta \right] =Z\left[ 0\right] \exp \left\{ -iT^2\sum_{n,m=-\infty
}^\infty \int \frac{d\mathbf{p}}{\left( 2\pi \right) ^3}\tilde{\eta}%
_n^{+}\left( \mathbf{p}\right) \gamma ^0\tilde{S}_F^{n-m}\left( \mathbf{p}%
\right) \tilde{\eta}_m\left( \mathbf{p}\right) \right\} ,
\end{equation}
where 
\begin{equation}
\int \frac{d\mathbf{p}}{\left( 2\pi \right) ^3}e^{i\mathbf{p\cdot x}}\tilde{S%
}_F^n\left( \mathbf{p}\right) =S_F^n\left( \mathbf{x}\right) .
\end{equation}
We may also write 
\begin{equation}
Z\left[ \eta \right] =Z\left[ 0\right] \exp \left\{ -iT^2\sum_{n,m=-\infty
}^\infty \int d\mathbf{x\,}d\mathbf{y}\eta _n^{+}\left( \mathbf{x}\right)
\gamma ^0S_F^{n-m}\left( \mathbf{x-y}\right) \eta _m\left( \mathbf{x}\right)
\right\} .
\end{equation}
where the propagators $S_F^n\left( \mathbf{x}\right) $ satisfy the equations 
\begin{equation}
\left\{ i\frac{U_n^{-1}-U_n}{2T}+\frac{_1}{^6}\overrightarrow{H}\left(
U_n+4+U_n^{-1}\right) \right\} S_F^n\left( \mathbf{x}\right) =-\frac{\delta
_n}T\gamma ^0\delta ^3\left( \mathbf{x}\right) .
\end{equation}
Using the rule 
\begin{equation}
\langle 0|\tilde{T}\tilde{\psi}_{na}^{+}\left( \mathbf{p}\right) \tilde{\psi}%
_{mb}\left( \mathbf{q}\right) |0\rangle =-i\left[ \tilde{S}_F^{m-n}\left( 
\mathbf{p}\right) \gamma ^0\right] _{ba}\left( 2\pi \right) ^3\delta
^3\left( \mathbf{p-q}\right)
\end{equation}
we find the following vacuum expectation values: 
\begin{eqnarray}
\langle 0|\tilde{\psi}_{n+1a}^{+}\left( \mathbf{p}\right) \tilde{\psi}%
_{nb}\left( \mathbf{q}\right) |0\rangle &=&\frac{18ie^{-i\theta }}{\left(
9+\kappa ^2\right) \sin \theta }\tilde{\Lambda}_{}^{+}\left( \mathbf{p}%
\right) _{ba}\left( 2\pi \right) ^3\delta ^3\left( \mathbf{p-q}\right) , 
\nonumber \\
\langle 0|\tilde{\psi}_{nb}^{+}\left( \mathbf{p}\right) \tilde{\psi}%
_{n+1a}\left( \mathbf{q}\right) |0\rangle &=&\frac{-18ie^{i\theta }}{\left(
9+\kappa ^2\right) \sin \theta }\tilde{\Lambda}_{}^{-}\left( \mathbf{p}%
\right) _{ab}\left( 2\pi \right) ^3\delta ^3\left( \mathbf{p-q}\right) , 
\nonumber \\
\langle 0|\tilde{\psi}_{n+1b}\left( \mathbf{p}\right) \tilde{\psi}%
_{na}^{+}\left( \mathbf{q}\right) |0\rangle &=&\frac{18ie^{-i\theta }}{%
\left( 9+\kappa ^2\right) \sin \theta }\tilde{\Lambda}_{}^{-}\left( \mathbf{p%
}\right) _{ba}\left( 2\pi \right) ^3\delta ^3\left( \mathbf{p-q}\right) , \\
\langle 0|\tilde{\psi}_{na}^{}\left( \mathbf{p}\right) \tilde{\psi}%
_{n+1b}^{+}\left( \mathbf{q}\right) |0\rangle &=&\frac{-18ie^{i\theta }}{%
\left( 9+\kappa ^2\right) \sin \theta }\tilde{\Lambda}^{+}\left( \mathbf{p}%
\right) _{ab}\left( 2\pi \right) ^3\delta ^3\left( \mathbf{p-q}\right) . 
\nonumber
\end{eqnarray}
We notice that these are singular at the parabolic barrier. However, by
taking anticommutators and we arrive at the fundamental quantisation
relations 
\begin{eqnarray}
\left\{ \tilde{\psi}_{n+1a}^{+}\left( \mathbf{p}\right) ,\tilde{\psi}%
_{nb}\left( \mathbf{q}\right) \right\} &=&\frac{36}{\left( 9+\kappa
^2\right) }\tilde{\Lambda}^{+}\left( \mathbf{p}\right) _{ba}\left( 2\pi
\right) ^3\delta ^3\left( \mathbf{p-q}\right) ,  \nonumber \\
\left\{ \tilde{\psi}_{nb}^{+}\left( \mathbf{p}\right) ,\tilde{\psi}%
_{n+1a}\left( \mathbf{q}\right) \right\} &=&\frac{36}{\left( 9+\kappa
^2\right) }\tilde{\Lambda}^{-}\left( \mathbf{p}\right) _{ab}\left( 2\pi
\right) ^3\delta ^3\left( \mathbf{p-q}\right) ,  \label{anticomm} \\
\left\{ \tilde{\psi}_{nb}^{+}\left( \mathbf{p}\right) ,\tilde{\psi}%
_{na}\left( \mathbf{q}\right) \right\} &=&0,  \nonumber
\end{eqnarray}
which are, remarkably, free of any singularities at the parabolic barrier.
If we had taken commutators instead, we would find that the singularities
still occurred at the parabolic barrier.

\subsection{Ladder operators}

Provided we are in the elliptic regime, the solution to the source free
Dirac equation 
\begin{equation}
\left( \hat{\Omega}^{+}\left( \mathbf{p}\right) \tilde{\psi}_{n+1}\left( 
\mathbf{p}\right) -2\eta _E\tilde{\psi}_n\left( \mathbf{p}\right) +\hat{%
\Omega}\left( \mathbf{p}\right) \tilde{\psi}_{n-1}\left( \mathbf{p}\right)
\right) \stackunder{c}{=}0
\end{equation}
is given by 
\begin{eqnarray}
\tilde{\psi}_n\left( \mathbf{p}\right) &=&\frac 1{2E}\sum_{r=1}^2\left\{
\left[ \hat{a}\left( \mathbf{p}r\right) e^{-in\delta }+\hat{c}\left( -%
\mathbf{p}r\right) e^{in\sigma }\right] u\left( \mathbf{p}r\right) \right. 
\nonumber \\
&&+\left. \left[ \hat{d}^{+}\left( \mathbf{p}r\right) e^{-in\sigma }+\hat{b}%
^{+}\left( -\mathbf{p}r\right) e^{in\delta }\right] v\left( \mathbf{-p}%
r\right) \right\} .
\end{eqnarray}
Then as outlined in the $1+0$ case, we find:

for the particles: 
\begin{eqnarray}
\hat{a}\left( \mathbf{p}r\right) &=&\frac i{2\sin \theta }u^{+}\left( 
\mathbf{p}r\right) \left\{ \tilde{\psi}_{n+1}\left( \mathbf{p}\right)
e^{-i\xi }-\tilde{\psi}_n\left( \mathbf{p}\right) e^{i\theta }\right\}
e^{in\delta },  \nonumber \\
\hat{a}^{+}\left( \mathbf{p}r\right) &=&\frac{-i}{2\sin \theta }\left\{ 
\tilde{\psi}_{n+1}^{+}\left( \mathbf{p}\right) e^{i\xi }-\tilde{\psi}%
_n^{+}\left( \mathbf{p}\right) e^{-i\theta }\right\} u\left( \mathbf{p}%
r\right) e^{-in\delta }, \\
\hat{b}\left( \mathbf{p}r\right) &=&\frac i{2\sin \theta }\left\{ \tilde{\psi%
}_{n+1}^{+}\left( -\mathbf{p}\right) e^{-i\xi }-\tilde{\psi}_n^{+}\left( -%
\mathbf{p}\right) e^{i\theta }\right\} v\left( \mathbf{p}r\right)
e^{in\delta },  \nonumber \\
\hat{b}^{+}\left( \mathbf{p}r\right) &=&\frac{-i}{2\sin \theta }v^{+}\left( 
\mathbf{p}r\right) \left\{ \tilde{\psi}_{n+1}\left( \mathbf{-p}\right)
e^{i\xi }-\tilde{\psi}_n\left( -\mathbf{p}\right) e^{-i\theta }\right\}
e^{-in\delta },
\end{eqnarray}

and for the oscillons: 
\begin{eqnarray}
\hat{c}\left( \mathbf{p}r\right) &=&\frac i{2\sin \theta }\left\{ \tilde{\psi%
}_{n+1}^{+}\left( -\mathbf{p}\right) e^{i\xi }-\tilde{\psi}_n^{+}\left( -%
\mathbf{p}\right) e^{i\theta }\right\} u\left( -\mathbf{p}r\right)
e^{in\sigma },  \nonumber \\
\hat{c}^{+}\left( \mathbf{p}r\right) &=&\frac{-i}{2\sin \theta }u^{+}\left( -%
\mathbf{p}r\right) \left\{ \tilde{\psi}_{n+1}\left( -\mathbf{p}\right)
e^{-i\xi }-\tilde{\psi}_n\left( -\mathbf{p}\right) e^{-i\theta }\right\}
e^{-in\sigma }, \\
\hat{d}\left( \mathbf{p}r\right) &=&\frac i{2\sin \theta }v^{+}\left( -%
\mathbf{p}r\right) \left\{ \tilde{\psi}_{n+1}\left( \mathbf{p}\right)
e^{i\xi }-\tilde{\psi}_n\left( \mathbf{p}\right) e^{i\theta }\right\}
e^{in\sigma },  \nonumber \\
\hat{d}^{+}\left( \mathbf{p}r\right) &=&\;\frac{-i}{2\sin \theta }\left\{ 
\tilde{\psi}_{n+1}^{+}\left( \mathbf{p}\right) e^{-i\xi }-\tilde{\psi}%
_n^{+}\left( \mathbf{p}\right) e^{-i\theta }\right\} v\left( -\mathbf{p}%
r\right) e^{-in\sigma }.
\end{eqnarray}
Here we have used conventional Dirac momentum spinors defined by 
\begin{eqnarray}
u\left( \mathbf{p}r\right) &\equiv &\frac{(\not{p}+m)}{\sqrt{m+E}}u_r=\frac
1{\sqrt{m+E}}\left[ 
\begin{array}{c}
\left( m+E\right) \chi _r \\ 
\mathbf{p\cdot \sigma }\chi _r
\end{array}
\right] ,  \nonumber \\
v\left( \mathbf{p}r\right) &\equiv &\frac{(m-\not{p})}{\sqrt{m+E}}v_r=\frac
1{\sqrt{m+E}}\left[ 
\begin{array}{c}
\mathbf{p\cdot \sigma }\eta _r \\ 
\left( m+E\right) \eta _r
\end{array}
\right]
\end{eqnarray}
with 
\begin{equation}
\hat{H}\left( \mathbf{p}\right) u\left( \mathbf{p}r\right) =u\left( \mathbf{p%
}r\right) ,\;\;\;\hat{H}\left( -\mathbf{p}\right) v\left( \mathbf{p}r\right)
=-v\left( \mathbf{p}r\right) .
\end{equation}
Now using the field anti-commutators, we arrive at the following non-zero
creation and annihilation relations: for the particles we have 
\begin{equation}
\left\{ \hat{a}\left( \mathbf{p}r\right) ,\hat{a}^{+}\left( \mathbf{q}%
s\right) \right\} =\left\{ \hat{b}\left( \mathbf{p}r\right) ,\hat{b}%
^{+}\left( \mathbf{q}s\right) \right\} =\frac{6E}{\sqrt{9-3\kappa ^2}}\delta
_{rs}\left( 2\pi \right) ^3\delta ^3\left( \mathbf{p-q}\right) ,
\end{equation}
whereas for the oscillons we find 
\begin{equation}
\left\{ \hat{c}\left( \mathbf{p}r\right) ,\hat{c}^{+}\left( \mathbf{q}%
s\right) \right\} =\left\{ \hat{d}\left( \mathbf{p}r\right) ,\hat{d}%
^{+}\left( \mathbf{q}s\right) \right\} =\frac{-6E}{\sqrt{9-3\kappa ^2}}%
\delta _{rs}\left( 2\pi \right) ^3\delta ^3\left( \mathbf{p-q}\right) .
\end{equation}
All other anticommutators are zero. It is clear we should be in the elliptic
regime for any of these anticommutators to make physical sense. Moreover, we
see that even though their linear momenta may be in the elliptic regime $TE_{%
\mathbf{p}}<\sqrt{3}$, oscillon and anti-oscillon particle states have a
negative inner product and are therefore unphysical. This confirms the
results of the previous section.

\subsection{Linear Momentum}

When there are continuous symmetries of the system function we may construct
various invariants of the motion using the \textit{Maeda-Noether }theorem
discussed in \textit{Principles I }\cite{J&N-I,MAEDA.81}.\textit{\ }First,
consider the case when there is invariance under translation in space. From
the results of \textit{Principles II }\cite{J&N-II} we construct the three
components of the conserved linear momentum using the rule 
\begin{eqnarray}
\mathbf{P}^n &\equiv &\int d\mathbf{x}\left\{ \left[ \mathcal{F}^n%
\overleftarrow{\frac \partial {\partial \psi _n}}-\nabla \mathbf{\cdot }%
\left( \mathcal{F}^n\overleftarrow{\frac \partial {\partial \nabla \psi _n}}%
\right) \right] \overrightarrow{\nabla }\psi _n\right.  \nonumber \\
&&\;\;\;\;\;\;\;\;\;\;\;\;\;\;\;\;\;\;\left. \;+\psi _n^{+}\overleftarrow{%
\nabla }\left[ \overrightarrow{\frac \partial {\partial \psi _n^{+}}}%
\mathcal{F}^n-\nabla \mathbf{\cdot }\left( \overrightarrow{\frac \partial
{\partial \nabla \psi _n^{+}}}\mathcal{F}^n\right) \right] \right\} .
\end{eqnarray}
This takes into account the anticommutation properties of the fields.
Applying this rule to the translation invariant source free system function
density $\left( \ref{fden}\right) $ we find 
\begin{equation}
\mathbf{P}^n=i\int d\mathbf{x}\left\{ \psi _n^{+}\overleftarrow{\nabla }%
\overrightarrow{\Lambda ^{+}}\psi _{n+1}-\psi _{n+1}^{+}\overleftarrow{%
\Lambda ^{-}}\overrightarrow{\nabla }\psi _n\right\}
\end{equation}
which has the appropriate limit 
\begin{equation}
\lim_{T\rightarrow 0}\mathbf{P}^n=\int d\mathbf{x}\left\{ \frac{_1}{^2}i\psi
^{+}\overleftarrow{\nabla }\psi -\frac{_1}{^2}i\psi ^{+}\overrightarrow{%
\nabla }\psi \right\} \;
\end{equation}
as expected. Using the equations of motion $\left( \ref{xeqm},\;\ref{yeqm}%
\right) $ we readily find 
\begin{equation}
\mathbf{P}^n\stackunder{c}{=}\mathbf{P}^{n+1}
\end{equation}
as expected. In the proof of this result we have to integrate by parts and
assume that the wave-function falls off to zero at spatial infinity, which
is to be expected for physical, i.e. normalisable states.

Taking Fourier transforms, we find 
\begin{equation}
\mathbf{P}^n=\int \frac{d\mathbf{p}}{\left( 2\pi \right) ^3}\mathbf{p}%
\left\{ \tilde{\psi}_n^{+}\left( \mathbf{p}\right) \tilde{\Lambda}^{+}\left( 
\mathbf{p}\right) \tilde{\psi}_{n+1}\left( \mathbf{p}\right) +\tilde{\psi}%
_{n+1}^{+}\left( \mathbf{p}\right) \tilde{\Lambda}^{-}\left( \mathbf{p}%
\right) \tilde{\psi}_n\left( \mathbf{p}\right) \right\}
\end{equation}
which on comparison with $\left( \ref{Logan}\right) $ shows that the linear
momentum is an example of a Logan invariant. After quantisation and using
the results $\left( \ref{anticomm}\right) $ we find the commutators 
\begin{eqnarray}
\left[ \mathbf{\hat{P}}^n,\hat{a}^{+}\left( \mathbf{p}r\right) \right] &=&%
\mathbf{p}\hat{a}^{+}\left( \mathbf{p}r\right) ,\;\;\;\left[ \mathbf{\hat{P}}%
^n,\hat{b}^{+}\left( \mathbf{p}r\right) \right] =\mathbf{p}\hat{b}^{+}\left( 
\mathbf{p}r\right) \\
\left[ \mathbf{\hat{P}}^n,\hat{c}^{+}\left( \mathbf{p}r\right) \right] &=&%
\mathbf{p}\hat{c}^{+}\left( \mathbf{p}r\right) ,\;\;\;\left[ \mathbf{\hat{P}}%
^n,\hat{d}^{+}\left( \mathbf{p}r\right) \right] =\mathbf{p}\hat{d}^{+}\left( 
\mathbf{p}r\right) ,
\end{eqnarray}
which shows that these operators do indeed create and annihilate excitations
carrying definite linear momentum.

\subsection{Angular momentum}

Consider the free Dirac system function $\left( \ref{fden}\right) $. An
infinitesimal rotation gives the following changes in the fields: 
\begin{equation}
\delta \psi _n=i\mathbf{\omega \cdot }\overrightarrow{\mathbf{J}}\psi
_n,\;\;\;\;\delta \psi _n^{+}=-i\psi _n\overleftarrow{\mathbf{J}}\mathbf{%
\cdot \omega },
\end{equation}
where 
\begin{equation}
\overrightarrow{\mathbf{J}}\equiv -i\mathbf{x\times \overrightarrow{\nabla }}%
+\frac{_1}{^2}\mathbf{\Sigma },\;\;\;\;\;\overleftarrow{\mathbf{J}}\equiv i%
\mathbf{x\times \overleftarrow{\nabla }}+\frac{_1}{^2}\mathbf{\Sigma }.
\end{equation}
The system function is invariant to this transformation and so we may use
the Maeda-Noether theorem to find the conserved angular momentum: 
\begin{equation}
\mathbf{J}_n=\int d\mathbf{x}\left\{ \psi _n^{+}\overleftarrow{\mathbf{J}}%
\overrightarrow{\Lambda ^{+}}\psi _{n+1}\right. +\psi _{n+1}^{+}\left. 
\overleftarrow{\Lambda ^{-}}\overrightarrow{\mathbf{J}}\psi _n\right\} .
\end{equation}
With the Cadzow equations$\;\left( \ref{xeqm},\;\ref{yeqm}\right) $ we find 
\begin{equation}
\mathbf{J}_n\stackunder{c}{=}\mathbf{J}_{n-1}
\end{equation}
as expected. We note in passing 
\begin{equation}
\left[ \overrightarrow{H},\overrightarrow{\mathbf{J}}\right] =\mathbf{0}.
\end{equation}
a relation encountered in the \textit{CT} Dirac equation.

\section{The charged Dirac equation}

We now discuss the coupling of electromagnetic potentials to the Dirac
field. We follow here the \textit{DT }formulation of the Maxwell fields
discussed in \textit{Principles II} $\cite{J&N-II},$ treating these fields
as external, i.e., non-dynamical, and so in this paper we consider only the
dynamics of the Dirac field. A full discussion of QED is reserved for
subsequent papers in this series.

In \textit{DT} gauge invariant electromagnetism, the electrostatic (scalar)
potential $\phi $ is associated with the \textit{temporal links} between
successive instants of time, whereas the magnetic (vector) potential $%
\mathbf{A}$ is associated with the sites themselves. If $\phi _n\left( 
\mathbf{x}\right) $ denotes the scalar potential at position $\mathbf{x}$ on
the temporal link $\left( n,n+1\right) $ and $\mathbf{A}_n\left( \mathbf{x}%
\right) $ is associated with the position $\mathbf{x}$ at time $n$, then
under a gauge transformation we have 
\begin{eqnarray}
\phi _n^{\prime }\left( \mathbf{x}\right) &=&\phi _n\left( \mathbf{x}\right)
+\frac{\chi _{n+1}\left( \mathbf{x}\right) -\chi _n\left( \mathbf{x}\right) }%
T,  \nonumber \\
\mathbf{A}_n^{\prime }\left( \mathbf{x}\right) &=&\mathbf{A}_n\left( \mathbf{%
x}\right) -\nabla \chi _n\left( \mathbf{x}\right) , \\
\psi _n^{\prime }\left( \mathbf{x}\right) &=&X_n\left( \mathbf{x}\right)
\psi _n\left( \mathbf{x}\right) ,  \nonumber
\end{eqnarray}
where $X_n\left( \mathbf{x}\right) \equiv \exp \left\{ -iq\chi _n\left( 
\mathbf{x}\right) \right\} ,$ $q$ being the charge associated with the Dirac
field.

Turning to virtual paths, the presence of electromagnetic potentials
requires a modification of the free field paths used previously. Following
our discussion of the charged Klein-Gordon field in \textit{Principles II},
we define the paths 
\begin{eqnarray}
\tilde{\psi}_n &\equiv &\lambda W_n^{\bar{\lambda}}\psi _{n+1}+\bar{\lambda}%
W_n^{-\lambda }\psi _n,  \nonumber \\
\tilde{\psi}_n^{+} &\equiv &\lambda W_n^{-\bar{\lambda}}\psi _{n+1}^{+}+\bar{%
\lambda}W_n^\lambda \psi _n^{+},
\end{eqnarray}
where $W_n\equiv \exp \left( iq\phi _nT\right) .$ The gauge covariant
derivatives are given by 
\begin{eqnarray}
\overrightarrow{D_n}\tilde{\psi}_n &\equiv &\left( \frac{\partial _\lambda }%
T+iq\phi _n\right) \tilde{\psi}_n=\frac{W_n^{\bar{\lambda}}\psi
_{n+1}-W_n^{-\lambda }\psi _n}T,  \nonumber \\
\overrightarrow{\mathbf{D}_n}\tilde{\psi}_n &\equiv &\left( \overrightarrow{%
\nabla }-iq\mathbf{\tilde{A}}_n\right) \tilde{\psi}_n \\
&=&\lambda W_n^{\bar{\lambda}}\left[ iqT\bar{\lambda}\nabla \phi _n+%
\overrightarrow{\nabla _n}-iq\mathbf{\tilde{A}}_n\right] \psi _{n+1}+\bar{%
\lambda}W_n^{-\lambda }\left[ -iqT\lambda \nabla \phi _n+\overrightarrow{%
\nabla _n}-iq\mathbf{\tilde{A}}_n\right] \psi _n,  \nonumber
\end{eqnarray}
where the virtual paths for the electromagnetic potentials are given by 
\begin{equation}
\tilde{\phi}_n\equiv \phi _n,\;\;\;\mathbf{\tilde{A}}_n\equiv \lambda 
\mathbf{A}_{n+1}+\bar{\lambda}\mathbf{A}_n.
\end{equation}
Then under a gauge transformation, we find 
\begin{eqnarray}
\overrightarrow{D_n^{\prime }}\tilde{\psi}_n^{\prime } &=&X_{n+1}^\lambda
X_n^{\bar{\lambda}}\overrightarrow{D_n}\tilde{\psi}_n=\tilde{X}_n%
\overrightarrow{D_n}\tilde{\psi}_n,  \nonumber \\
\overrightarrow{\mathbf{D}_n^{\prime }}\tilde{\psi}_n^{\prime }
&=&X_{n+1}^\lambda X_n^{\bar{\lambda}}\overrightarrow{\mathbf{D}_n}\tilde{%
\psi}_n=\tilde{X}_n\overrightarrow{\mathbf{D}_n}\tilde{\psi}_n.
\end{eqnarray}

The gauge invariant system function for the Dirac field in external
electromagnetic potentials may be given in the form 
\begin{equation}
F^n=\frac{_1}{^2}T\int d\mathbf{x}\left\{ \langle \tilde{\psi}_n^{+}%
\overrightarrow{\mathcal{O}_n}\tilde{\psi}_n\rangle +\langle \tilde{\psi}%
_n^{+}\overleftarrow{\mathcal{O}_n}\tilde{\psi}_n\rangle \right\}
\end{equation}
where we use the angular brackets to denote integration over $\lambda $,
i.e. 
\begin{equation}
\langle f\rangle \equiv \int_0^1f\left( \lambda \right) d\lambda
\end{equation}
and 
\begin{eqnarray}
\overrightarrow{\mathcal{O}_n} &\equiv &i\overrightarrow{D_n}+i\mathbf{%
\alpha \cdot }\overrightarrow{\mathbf{D}_n}-m\beta ,  \nonumber \\
\overleftarrow{\mathcal{O}_n} &\equiv &-i\overleftarrow{D_n}-i\mathbf{\alpha
\cdot }\overleftarrow{\mathbf{D}_n}-m\beta .
\end{eqnarray}
Then we find the Cadzow equations of motion for the Dirac fields are 
\begin{eqnarray}
&&\langle \bar{\lambda}W_n^\lambda \overrightarrow{\mathcal{O}_n}\tilde{\psi}%
_n\rangle +\langle \lambda W_{n-1}^{-\bar{\lambda}}\overrightarrow{\mathcal{O%
}_{n-1}}\tilde{\psi}_{n-1}\rangle \stackunder{c}{=}0,  \nonumber \\
&&\langle \tilde{\psi}_n^{+}\overleftarrow{\mathcal{O}_n}\bar{\lambda}%
W_n^{-\lambda }\rangle +\langle \tilde{\psi}_{n-1}^{+}\overleftarrow{%
\mathcal{O}_{n-1}}\lambda W_{n-1}^{\bar{\lambda}}\rangle \stackunder{c}{=}0.
\end{eqnarray}

We note that, as found in \textit{Principles III}, there is often an
advantage in not evaluating the integration over the virtual path until a
late stage in a calculation.

To go further, we introduce the following notation 
\begin{eqnarray}
A_n &\equiv &\lambda W_{n-1}^{\bar{\lambda}}\psi _n,\;\;\;B_n\equiv \bar{%
\lambda}W_n^{-\lambda }\psi _n  \nonumber \\
A_n^{+} &\equiv &\lambda W_{n-1}^{-\bar{\lambda}}\psi
_n^{+},\;\;\;B_n^{+}\equiv \bar{\lambda}W_n^\lambda \psi _n^{+}.
\end{eqnarray}
Then we have 
\begin{equation}
\tilde{\psi}_n=A_{n+1}+B_n,\;\;\;\tilde{\psi}_n^{+}=A_{n+1}^{+}+B_n^{+}
\end{equation}
and then the Cadzow equations give the relations 
\begin{eqnarray}
&&\langle B_n^{+}\overrightarrow{\mathcal{O}_n}\left( A_{n+1}+B_n\right)
\rangle +\langle A_n^{+}\overrightarrow{\mathcal{O}_{n-1}}\left(
A_n+B_{n-1}\right) \rangle \stackunder{c}{=}0  \nonumber \\
&&\langle \left( A_{n+1}^{+}+B_n^{+}\right) \overleftarrow{\mathcal{O}_n}%
B_n\rangle +\langle \left( A_n^{+}+B_{n-1}^{+}\right) \overleftarrow{%
\mathcal{O}_{n-1}}A_n\rangle \stackunder{c}{=}0.  \label{ceqn}
\end{eqnarray}
These relations are very useful for proving charge conservation. Under gauge
transformations where 
\begin{equation}
W_n^{\prime }=X_{n+1}^{-1}W_nX_n,\;\;\;X_n\equiv \exp \left\{ -iq\chi
_n\right\} ,\;\;\;\psi _n^{\prime }=X_n\psi _n,
\end{equation}
then 
\begin{equation}
A_n^{\prime }=\tilde{X}_{n-1}A_n,\;\;\;B_n^{\prime }=\tilde{X}_nB_{n,}\;\;\;%
\tilde{\psi}_n^{\prime }=\tilde{X}_n\tilde{\psi}_n
\end{equation}
where 
\begin{equation}
\tilde{X}_n=\exp \left\{ -iq(\lambda \chi _{n+1}+\bar{\lambda}\chi
_n\right\} =X_{n+1}^\lambda X_n^{\bar{\lambda}}.
\end{equation}
To find the total charge consider the infinitesimal global gauge
transformation 
\begin{equation}
\delta \psi _n=-iq\delta \chi \psi _n,\;\;\;\delta \psi _n^{+}=iq\delta \chi
\psi _n^{+}
\end{equation}
and apply the Maeda-Noether theorem discussed in \textit{Principles II}.
Then we find the total charge can be written in the form 
\begin{equation}
Q^n=iqT\int d\mathbf{x}\left\{ \langle A_{n+1}^{+}\overleftarrow{\mathcal{O}%
_n}B_n\rangle -\langle B_n^{+}\overrightarrow{\mathcal{O}_n}A_{n+1}\rangle
\right\}
\end{equation}
Then modulo the equations of motion $\left( \ref{ceqn}\right) $ we readily
find 
\begin{equation}
Q^n\stackunder{c}{=}Q^{n-1}
\end{equation}
and 
\begin{equation}
\lim_{T\rightarrow 0}Q^n=q\int d\mathbf{x}\psi ^{+}\psi
\end{equation}
as expected.

\subsection{Reduction formulae}

We anticipate now the construction of scattering amplitudes to be discussed
in detail in subsequent papers of this series by giving the reduction
formulae involving the Dirac particle and anti-particle creation and
annihilation operators. If $\tilde{T}$ denotes the discrete time ordering
operator, discussed above and $\overrightarrow{U_{n\text{ }}}$ is the
classical step function defined previously then we find the following
reduced matrix elements:

\begin{eqnarray}
\langle \alpha _{out}|a_{out}\left( \mathbf{p}r\right) \left( \tilde{T}\hat{O%
}\right) |\beta _{in}\rangle _R &=&\frac i{2\sin \theta }\sum_{n=-\infty
}^\infty e^{in\delta }u^{+}\left( \mathbf{p}r\right) \overrightarrow{%
\mathcal{D}\left( \mathbf{p}\right) }\langle \alpha _{out}|\tilde{T}\left(
\psi _n\left( \mathbf{p}\right) \hat{O}\right) |\beta _{in}\rangle , 
\nonumber \\
\langle \alpha _{out}|(\tilde{T}\hat{O})a_{in}^{+}\left( \mathbf{p}r\right)
|\beta _{in}\rangle _R &=&\frac i{2\sin \theta }\sum_{n=-\infty }^\infty
\langle \alpha _{out}|\tilde{T}\left( \hat{O}\psi _n^{+}\left( \mathbf{p}%
\right) \right) |\beta _{in}\rangle \overleftarrow{\mathcal{D}\left( \mathbf{%
p}\right) }u\left( \mathbf{p}r\right) e^{-in\delta },  \nonumber \\
&& \\
\langle \alpha _{out}|b_{out}\left( \mathbf{p}r\right) \left( \tilde{T}\hat{O%
}\right) |\beta _{in}\rangle _R &=&\frac i{2\sin \theta }\sum_{n=-\infty
}^\infty \langle \alpha _{out}|\tilde{T}\left( \psi _n^{+}\left( -\mathbf{p}%
\right) \hat{O}\right) |\beta _{in}\rangle \overleftarrow{\mathcal{D}\left( -%
\mathbf{p}\right) }v\left( \mathbf{p}r\right) e^{in\delta },  \nonumber \\
\langle \alpha _{out}|(\tilde{T}\hat{O})b_{in}^{+}\left( \mathbf{p}r\right)
|\beta _{in}\rangle _R &=&\frac i{2\sin \theta }\sum_{n=-\infty }^\infty
e^{-in\delta }v^{+}\left( \mathbf{p}r\right) \overrightarrow{\mathcal{D}%
\left( -\mathbf{p}\right) }\langle \alpha _{out}|\tilde{T}\left( \hat{O}\psi
_n\left( -\mathbf{p}\right) \right) |\beta _{in}\rangle  \nonumber \\
&&
\end{eqnarray}
where 
\begin{eqnarray}
\overrightarrow{\mathcal{D}\left( \mathbf{p}\right) } &\equiv &\hat{\Omega}%
^{+}\left( \mathbf{p}\right) \overrightarrow{U_n}-2\eta _{\mathbf{p}}+\hat{%
\Omega}\left( \mathbf{p}\right) \overrightarrow{U_n^{-1}},  \nonumber \\
\overleftarrow{\mathcal{D}\left( \mathbf{p}\right) } &\equiv &\overleftarrow{%
U_n}\hat{\Omega}\left( \mathbf{p}\right) -2\eta _{\mathbf{p}}+\overleftarrow{%
U_n^{-1}}\hat{\Omega}^{+}\left( \mathbf{p}\right) .
\end{eqnarray}
In these matrix elements we have discarded the non-scattered components, as
usual. In addition, we have not given any reduction formulae for the
oscillon and anti-oscillon ladder operators. These we regard as not creating
physically accessible states, and therefore, they should not be put on the
same footing as the particle and anti-particle creation and annihilation
operators.

\section{Concluding remarks}

The introduction of a non-zero $T$ opens up a crack in \textit{CT} quantum
field theory which admits a number of new phenomena. In addition to the
novelties encountered previously in \textit{DT }bosonic theories, such as a
physical particle momentum cutoff, modified propagators and vertices, we
find in the case of the Dirac equation the appearance of fermionic \textit{%
oscillon }and \textit{anti-oscillon} solutions. These should not survive in
the \textit{CT} limit on account of their extraordinary properties. For
non-zero $T$ they should certainly participate in \textit{DT} Feynman
diagram processes as intermediate, virtual objects carrying charge, linear
momentum and spin, but we expect them not to be observable as asymptotic 
\textit{in} or \textit{out} particles.\ We plan to investigate their role in 
\textit{DT} QED scattering processes and regularisation in subsequent papers
in this series.

\section{Acknowledgement}

Keith Norton is grateful to the Crowther Fund of the Open University for
financial assistance during this course of this research.

\end{document}